\documentclass{aa}
\usepackage{graphicx}
\usepackage{amsmath}

\begin{document}

   \title{Spiral density wave generation by vortices in Keplerian flows}

\author{G. Bodo\inst{1} \and G. Chagelishvili\inst{3} \and G. Murante\inst{1}
  \and A. Tevzadze\inst{3} \and P. Rossi \inst{1} \and A. Ferrari\inst{2}
            }

   \offprints{G. Bodo}

   \institute{
INAF Osservatorio Astronomico di Torino, Strada dell'Osservatorio
             20, I-10025 Pino Torinese, Italy\\
             email: bodo@to.astro.it
\and
Dipartimento di Fisica Generale dell'Universit\`a,
Via Pietro Giuria 1, I - 10125 Torino, Italy
\and
Center for Plasma Astrophysics, Abastumani
Astrophysical Observatory, 2a Kazbegi Ave. Tbilisi 0160, Georgia                              }

   \date{Received; accepted }

\abstract{
We perform a detailed analytical and numerical study of the dynamics
of perturbations (vortex/aperiodic mode, Rossby and spiral-density
waves) in 2D compressible disks with a Keplerian law of rotation. We
draw  attention to the process of spiral-density wave generation
from vortices, discussing, in particular, the initial, most peculiar
stages of  wave emission. We show that the linear phenomenon of
wave generation by vortices in smooth (without inflection points)
shear flows found by using the
so-called non-modal approach, is directly applicable to the present
case. After an analytical non-modal description of the physics and
characteristics of the spiral-density wave generation/propagation in
the  local shearing-sheet model, we follow 
the process of wave generation by small amplitude coherent circular
vortex structures, by direct  global 
numerical simulation, describing the main features of the generated
waves.   }

\maketitle

\section{Introduction}

The dynamics of vortices in astrophysical disks has recently
received much interest both because vortices in protoplanetary
disks can represent aggregation regions of solid particles for the
eventual formation of planets (Barge \& Sommeria 1995) and more
generally for understanding  accretion disk dynamics and the
basic problem of angular momentum transport (Lovelace et al. 1999,
Li et al. 2000).  Several works have been devoted to the analysis
of the possibility of forming and maintaining coherent vortex
structures in the strongly sheared flow pertaining to a Keplerian
disk, both in   barotropic configurations,
where the initial potential vorticity perturbation is conserved
(Bracco et al. 1999, Godon \& Livio 1999,2000, Davis et al.
2000, Davis 2002) and in baroclinic situations where  one can have
vorticity generation (Klahr \& Bodenheimer 2003, Klahr 2004).
 In the incompressible case it has been shown
that coherent vortex structures can indeed form (under conservation
of potential vorticity) and anticyclonic vortices can survive longer
than cyclonic ones (Bracco et al. 1999) and give rise to the
appearance of  Rossby waves in the system (Davis et al. 2000). The
effects of compressibility have not yet been fully analyzed and
require  rigorous study. Godon \& Livio (1999)
performed two-dimensional time-dependent numerical simulations of
vortices in  viscous compressible Keplerian disks. 
Vorticity waves are considered as one of the  constituents of
(anticyclonic) vortex dynamics, but without specification of the
wave properties and any analysis of their genesis and dynamics
(the subject of the study was the stability and lifetime of vortices).
Davis (2002) performed fully
compressible numerical simulations of the dynamics of a single
vortical structure in a Keplerian disk flow and has reported the
generation of outward-moving compressible waves by the coherent
vortex (the generation was attributed to nonlinear processes)
pointing out the potential importance of this phenomenon for 
vortex dynamics. Johansen, Andersen \& Brandenburg
(2004) considered the dynamics of nonlinear vortices by numerical
3D simulations in the local shearing sheet approximation, also observing
indications of wave generation. Klahr \& Bodenheimer
(2003), performing 2D and 3D hydrodynamical simulations of
protoplanetary disks,  found that a radial entropy gradient can
generate Rossby waves which eventually break into vortices. 
Klahr (2004), by a linear stability
analysis similar to the one presented in this paper, in fact 
showed that a radial
entropy gradient leads to continuous generation of potential
vorticity and to a  transient swing-like amplification of the
vortical/aperiodic mode, without the subsequent decay observed in
the linear barotropic configuration.

The aim of this paper is to investigate the linear dynamics of
initially-imposed vortical perturbations in 2D compressible
Keplerian disks to understand the phenomena of spiral-density and
Rossby wave generation
 and to study the initial, most peculiar stages
of wave emission/propagation. In this repect, we want to stress that the origins 
of spiral-density and Rossby waves are absolutely different.

We base our understanding of spiral-density wave generation on the
papers by Chagelishvili et al. (1997, 2000), where 
{\it a completely linear phenomenon of acoustic wave
generation by vortex mode perturbation in smooth shear flows}
was for the first time discussed. It was
also shown that the generated acoustic waves are emitted as two
symmetric packages in opposite directions from the parent
coherent vortex perturbation.  This wave generation
phenomenon is directly applicable to the generation of spiral-density
waves by vortices in  disk flows (Sec. 3). The Rossby wave appearance
from the vortex mode perturbations, on the contrary, is not connected to
any generation process. The essence of the Rossby wave formation lies 
in the ``breaking of degeneracy'' of vortex mode perturbations
(Sec. 4).

Recently, much progress has been made by the hydrodynamic
community in the analysis of the linear dynamics of  smooth shear
flows by recognising that the traditional normal mode
approach does not represent the best tool for its study because of the
non-normality of the linear operators in the flows (cf. Reddy, Schmid
\& Henningson 1993, Henningson \& Reddy 1994, Gustavsson 1991, Farrell
\& Ioannou 1993, Craik \& Criminale 1986). This means that the
corresponding eigenfunctions are not orthogonal and strongly interfere.
The knowledge drawn from the analysis of separate eigenfunctions
and eigenvalues is far from complete. A correct and full
description of the shear flow phenomena needs knowledge of the
interference processes which cannot be easily taken into account in
the framework of modal analysis. These circumstances led to the
development of the so-called non-modal approach that could reveal several
unexpected phenomena, which were overlooked by the normal mode
analysis. The non-modal analysis is a modification of the initial
value problem. It involves the change of independent variables from
the laboratory to a moving frame and the study of the temporal evolution of
spatial Fourier harmonics (SFH) of perturbations without any spectral
expansion in time. The resulting linear dynamics is much richer than
that expected from the stability predictions of the modal approach and
these linear processes may also play a fundamental role in the full
dynamics of smooth shear flows.  For example, it has been shown that
vortical perturbations can have a phase of transient amplification,
which is at the base of the so-called ``bypass'' concept of the onset
of turbulence in planar Couette flows (Baggett et al 1995, Gebhardt \&
Grossman 1994, Farrell \& Ioannou 1993, Henningson \& Ready 1994,
Grossmann 2000, Chagelishvili et al 2002, Chapman 2002). Another
phenomenon of interest here is the generation of waves by
a vortex/aperiodic mode of perturbations, which has been discussed 
for the first time, as
we saw above, by Chagelishvili et al. (1997) and analyzed by
Chagelishvili et al. (2000) and Farrell \& Ioannou (2000) in the 
simplest planar configuration. In this paper we extend the non-modal
analysis of this process to the case of 2D Keplerian disk flow (that
is a natural example of smooth shear flows) in the shearing
sheet approximation, giving the theoretical basis to understand
the results of simulations like those presented by Davis (2002). We
also perform numerical simulations of the vortex dynamics in Keplerian
disks: we consider a small amplitude (to include only linear processes),
coherent circular vortex and focus on wave
generation and emission, on the properties of density and vorticity
fields and on the description of the wave propagation
trajectories. Physical model and equations are presented in
Sec. 2. The non-modal analysis of spiral-density wave
generation/emission by vortex mode perturbation is presented in
Sec. 3. The appearance of the  Rossby wave from vortex mode perturbation is
outlined in Sec. 4. Numerical simulations of  wave emergence from
a coherent circular vortex perturbation is presented in Sec. 5 and
in the last section we summarize our results. In the Appendix we
describe the  method for the selection of initial values of pure vortex
mode perturbations for our numerical study.

\section{Physical model and equations}
\label{sec:theory}

In cylindrical coordinates, the basic equations will
read as follows:
\begin{equation}
{\partial \rho \over \partial t} + {\partial \over \partial
r}(\rho V_r) + {1 \over r} {\partial \over \partial \phi} (\rho
V_\phi) = 0 ~,
\label{hydro1}
\end{equation}
\begin{equation}
{\partial V_r \over \partial t} + ({\bf V} \nabla) V_r - {V_\phi^2
\over r}= - {1 \over \rho}{\partial P \over \partial r} -
{\partial \Phi \over \partial r}~,
\label{hydro2}
\end{equation}
\begin{equation}
{\partial V_\phi \over \partial t} + ({\bf V} \nabla) V_\phi +
{V_r V_\phi \over r} = - {1 \over \rho r}{\partial P \over
\partial \phi}~,
\label{hydro3}
\end{equation}
\begin{equation}
\left({\partial \over \partial t} + ({\bf V} \nabla) \right) P =
\gamma {P \over \rho
} \left({\partial \over \partial t} + ({\bf V}
\nabla) \right) \rho~,
\label{hydro4}
\end{equation}
where
\begin{equation}
({\bf V} \nabla) \equiv V_r {\partial \over \partial r} + {V_\phi
\over r} {\partial \over \partial \phi} ~.
\label{defvgradv}
\end{equation}

We consider a two dimensional (2D) inviscid, differentially
rotating flow. The flow has axial symmetry and a rotation axis parallel
to the $~z~$ axis: $~{\bf V_0} = (0,V_{0\phi},0)$, with
$~V_{0\phi}=r\Omega({\bf r})$. In this flow configuration the basic
force balance is described by:
\begin{equation}
r \Omega^2(r) = {1 \over \rho_0} {\partial P_0 \over \partial r} +
{\partial \Phi({\bf r}) \over \partial r},
\label{forcebalance}
\end{equation}
We choose a distribution of angular velocity that follows the
Keplerian law
\begin{equation}
\Omega^2({\bf r}) \propto {1 \over r^3},
\label{kepl}
\end{equation}
and a gravitational potential that balances the centrifugal term
\begin{equation}
\Phi({\bf r})  \propto -{1 \over r},
\label{gravity}
\end{equation}
the pressure is then constant and equal to $P_0$.

Our 2D model of a Keplerian disk retains the
dynamical effects of differential rotation, while  discarding geometrical
and thermodynamic $~(~ P_0,~ \rho_0 = const)~$ complications. The sound
speed is defined by:
\begin{equation}
c_s^2 = {\gamma P_0 \over \rho_0} = {\rm const}.
\label{sound}
\end{equation}
with the specific heat ratio $~\gamma = 5/3 ~$. Even though our flow
is two-dimensional and has no $z$ dependence, it is useful for the
discussion to introduce  the
parameter
\begin{equation}
H = c_s/\Omega \quad,
\label{thickness}
\end{equation}
that represents the equivalent thickness of a thin disk model.

In a Keplerian disk the basic flow vorticity is (cf. Tagger
2001):
\begin{equation}
W = {1 \over 2\Omega}\left[4\Omega^2 +2\Omega \left( r{\partial
\Omega \over \partial r}\right) \right]={\Omega \over 2}.
\label{vorticity}
\end{equation}
Thus, the mean vorticity gradient across the disk is nonzero and
should not a priori be neglected.

\section{ Linear dynamics of vortex mode -- Generation of spiral-density waves}

Here we study the linear dynamics of 2D small-scale perturbations
(i.e. with characteristic scales in radial and
azimuthal directions much less than the radial characteristic
scale of the background/disk flow) in the shearing sheet
approximation (e. g., Goldreich \& Lynden-Bell 1965; Goldreich \&
Tremaine 1978; Nakagawa \& Sekiya 1992). In this case the dynamical
equations are written in the local co-moving Cartesian co-ordinate
system:
\begin{equation}
x \equiv r - r_0; ~~~~~ y \equiv r_0 (\phi - \Omega_0 t); ~~~~~{x
\over r_0} ,~ {y \over r_0} \ll 1,
\label{shearingsheet1}
\end{equation}
where ($r, \phi$) are standard cylindrical co-ordinates and
$\Omega_0$ is the local rotation angular velocity at $r=r_0$:
\begin{equation}
 \Omega(r)= \Omega_0 +  {\partial  \Omega  \over \partial r} (r-r_0)\equiv
\Omega_0 +  2A {x \over r_0}, ~~~~~
\label{shearingsheet2}
\end{equation}
where
\begin{equation}
A \equiv {1 \over 2} \left[r {\partial \Omega \over \partial r}
\right]_{r=r_0}
\label{Oort1}
\end{equation}
is the standard Oort constant which describes the mean velocity
shear parameter in the local frame. In the case of a Keplerian
rotation law:
\begin{equation}
A = -{3 \over 4} \Omega_0 < 0.
\label{Oort2}
\end{equation}
The main assumption of the shearing sheet approximation is the
neglect of the basic flow vorticity gradient; the only gradient
being considered is  differential rotation.

Introducing linear perturbations
\begin{equation}
{\bf V} = {\bf V}_0 + {\bf v}^\prime, ~~~ P = P_0 + p^\prime, ~~~
\rho = \rho_0 + \rho^\prime,
\label{linearize}
\end{equation}
 we derive, from system \ref{hydro1}-\ref{hydro4}, the equations
governing the linear
dynamics of perturbations in the local Cartesian frame:
\begin{equation}
\left({\partial \over \partial t} + 2Ax {\partial \over \partial
y}\right) v_x^\prime - 2\Omega_0 v_y^\prime + {1 \over
\rho_0}{\partial p^\prime \over \partial x} = 0,
\label{hydro-linear1}
\end{equation}
\begin{equation}
\left({\partial \over \partial t} + 2Ax {\partial \over \partial
y}\right) v_y^\prime + 2(\Omega_0 + A) v_x^\prime + {1 \over
\rho_0}{\partial p^\prime \over \partial y} = 0,
\label{hydro-linear2}
\end{equation}
\begin{equation}
\left({\partial \over \partial t} + 2Ax {\partial \over \partial
y}\right) \rho^\prime + {\rho_0}\left({\partial v_x^\prime \over
\partial x} + {\partial v_y^\prime \over \partial y}\right) = 0,
\label{hydro-linear3}
\end{equation}
\begin{equation}
\left({\partial \over \partial t} + 2Ax {\partial \over \partial
y}\right) p^\prime + c_{\rm s}^2\left({\partial \over \partial t}
+ 2Ax {\partial \over \partial y}\right) \rho^\prime = 0.
\label{hydro-linear4}
\end{equation}

Hence, following the standard method of non-modal analysis
(cf. Goldreich \& Lynden-Bell 1965; Goldreich \& Tremaine 1978),
we introduce the spatial Fourier harmonics (SFH) of
perturbations with time-dependent phases:
\begin{equation}
\left\{\begin{array}{c} {v_x}^\prime({\bf r},t)\\
{v_y}^\prime({\bf r},t)\\ p^\prime({\bf r},t) \\ \rho^\prime({\bf
r},t)\\
\end{array}\right\} = \\
\left\{\begin{array}{c} v_x(k_x(t),k_y,t)\\ v_y(k_x(t),k_y,t)\\
p(k_x(t),k_y,t) \\ \varrho(k_x(t),k_y,t)
\end{array}\right\}
\exp ( {\rm i}k_x(t)x + {\rm i}k_yy)
\label{fourier}
\end{equation}
\begin{equation}
 k_x(t) = k_x(0) - 2A k_y t=k_x(0) +\frac{3}{2}\Omega_0 k_y t.
\label{drift}
\end{equation}

The streamwise/azimuthal wavenumber remains constant and the
streamcross/radial wavenumber changes at a constant rate; in
the linear approximation SFH ``drifts'' in the $~{\bf k}$-plane
(wavenumber plane). The effect of this change is that lines of
constant phase (wave crests) are rotated by the basic flow.

Substitution of Eqs. (\ref{fourier},\ref{drift}) into
Eqs. (\ref{hydro-linear1}-\ref{hydro-linear4})  yields the system of
ordinary differential equations that governs  the linear dynamics of
SFH of perturbations in the described flow:
\begin{equation}
{{d  v_x(t)} \over {d t}} = 2\Omega_0 v_y(t) - {\rm i} k_x(t)
{p(t)\over \rho_0},
\label{eqvx}
\end{equation}
\begin{equation}
{{d v_y(t)} \over {d t}}  = -2(\Omega_0 + A) v_x(t) - {\rm i} k_y
{p(t)\over \rho_0},
\label{eqvy}
\end{equation}
\begin{equation}
{{d \varrho(t)} \over {d t}} +{\rm i} \rho_0 \left[ k_x(t) v_x(t)
+ k_y v_y(t) \right] =0,
\label{eqrho}
\end{equation}
\begin{equation}
p(t)= c_{\rm s}^2 \varrho(t).
\label{eqpr}
\end{equation}

Introducing $~D(t) \equiv {{\rm i} \varrho(t)/\rho_0} $, Eqs.
(23-26) become:
\begin{equation}
{{d v_x(t)} \over {d t}} = 2\Omega_0 v_y(t) - k_x(t) c_{\rm s}^2
D(t),
\label{eqvx2}
\end{equation}
\begin{equation}
{{d v_y(t)} \over {d t}} = -2(\Omega_0 + A) v_x(t) - k_y c_{\rm
s}^2 D(t),
\label{eqvy2}
\end{equation}
\begin{equation}
{{d D(t)} \over {d t}} = k_x(t) v_x(t) + k_y v_y(t).
\label{eqrho2}
\end{equation}
Klahr (2004)  obtained a similar system of equations for a
configuration with a radial entropy gradient in the basic state, in
that case he gets a fourth order system, while in our case, as we
will see, the system can be reduced to second order.

The latter system is characterized by the important time
invariant:
\begin{equation} k_x(t) v_y(t) - k_y v_x(t) + 2(\Omega_0 + A) D(t) \equiv {\cal
I}
\label{invariant}
\end{equation}
that follows (for SFH of perturbations) from the conservation of
potential vorticity. This time invariant, in turn, indicates the
existence of the vortex/aperiodic mode in the perturbation spectrum
whose dynamics represent the main interest of this Section.  A
radial entropy gradient would introduce a source term for potential
vorticity (Klahr 2004) and the quantity above  would no longer be invariant.

We can define the spectral density of the total energy in the ${~\bf
k}$-plane as:
\begin{equation}
E_{\bf k}(t) \equiv {\rho_0 \over 2} \left (v_x^2+v_y^2
\right)+{\rho_0c_s^2 \over 2} D^2,
\label{energy}
\end{equation}
where the two terms correspond to the kinetic and
potential energies of SFH.

The numerical study of SFH dynamics is governed by
Eqs. (\ref{eqvx2}-\ref{invariant}).
However, for the fundamental comprehension of the physical nature
of the perturbations and their linear dynamics in the flow,  we
 rewrite them in another form: from
Eqs. (\ref{eqvx2}-\ref{invariant})  one
can get the following second order {\it inhomogeneous}
differential equation for $~v_y(t)$:
\begin{equation}
{{d^2 v_y(t)} \over {d t^2}} + \omega_{SD}^2(t)v_y(t)=
k_x(t)c_{\rm s}^2 {\cal I},
\label{secondorder}
\end{equation}
and the equations expressing $~v_x(t)~$ and $~D(t)$ as functions of
$~v_y(t)~$ and its time derivative:
\begin{multline}
v_x(t) = {1 \over {4(\Omega_0 + A)^2 + k_y^2 c_s^2}} \times \\
\times \left[ k_x(t)
k_y c_s^2 v_y(t) - 2 (\Omega_0 + A){{d v_y(t)} \over {d t}} -  k_y
c_s^2{\cal I} \right],
\label{defvx}
\end{multline}
\begin{equation}
D(t) = {1 \over 2 (\Omega_0 + A)} \left[ k_x(t) v_y(t) - k_y
v_x(t) - {\cal I} \right],
\label{defD}
\end{equation}
where
\begin{equation}
~\omega_{SD}^2(t) \equiv 4\Omega_0(\Omega_0
+ A) + k^2(t) c_s^2 = \Omega_0^2 + k^2(t) c_s^2 ~,
\label{disprel}
\end{equation}
and $~k^2(t) \equiv k_x^2(t) + k_y^2$.

Eq. \ref{secondorder} describes two different modes of
perturbations:\\
{\bf(a)} Spiral-density wave mode ${~(~v_y^{\rm (w)})}$, that is described
by the general solution of the corresponding homogeneous
equation. Notice that the frequency of the wave $~\omega(t)~$ is time-dependent.\\
{\bf (b)} Vortex mode ${~(~v_y^{\rm (v)})}$, that is
aperiodic, originates from the equation inhomogeneity
${~(~k_x(t)c_s^2{\cal I})~}$ and is associated with the particular
solution of the inhomogeneous equation. The amplitude of the vortex
mode is proportional to ${~{\cal I}}$ and goes to zero when ${~{\cal
I}} = 0$. The existence of the vortex/aperiodic mode, as discussed in
more detail in Section 4, is the result of the main assumption of the
shearing sheet approximation, i.e. the neglect of the basic flow
vorticity gradient. Also, the vortex mode acquires nonzero divergence
and nonzero density perturbation ($~D^{\rm (v)} \neq 0$, where
$~D^{\rm (v)}$ is the density perturbation of the vortex mode as
defined below).

Thus, we can decompose any perturbation as the sum of the wave and
vortex mode perturbations:
\begin{equation}
v_{x} = v_{x}^{\rm (w)} + v_{x}^{\rm (v)};~~~ v_{y} = v_{y}^{\rm
(w)} + v_{y}^{\rm (v)};~~~D = D^{\rm (w)} + D^{\rm (v)}.
\label{decomposition}
\end{equation}

 Clearly, the character of the
dynamics depends on which mode of perturbation is imposed initially
in Eqs. (\ref{eqvx2}-\ref{eqrho2}): pure waves (without mixture of aperiodic
vortices), pure aperiodic vortices (without mixture of waves) or a
mixture of the two.

Here we present a numerical and qualitative study of the linear
dynamics of an initially excited SFH of vortex mode perturbation.
The dynamics is defined by the value of the parameter:
\begin{multline}
R(t) \equiv {2|A| \over {\sqrt {4\Omega_0(\Omega_0 + A) +
[k_x^2(t) + k_y^2] c_{\rm s}^2}}} = \\
= {3 \over 2\sqrt {1+[k_x^2(t) +
k_y^2]H^2 }}~,
\end{multline}
where $H$ is the equivalent thickness of a thin disk model defined in
Eq. (\ref{thickness}).
 $R(t)$ is a time-dependent parameter and
consequently, the character of the dynamics is also
time-dependent. The final result of the dynamics is defined by the
maximum value of $~R(t)~$ (achieved at $~t=t^*$, when
$~k_x(t^*)=0$):
\begin{equation}
R_{max} = {3 \over 2\sqrt {1+ k_y^2H^2 }}\equiv {3 \over 2\sqrt {1+
(m H / R)^2 }}~,
\end{equation}
 where $m$ is the azimuthal wave-number of perturbation SFH.

We carried out numerical integration of
Eqs. (\ref{eqvx2}-\ref{eqrho2})  for an
initially imposed leading vortex mode SFH ($~k_x(0) / k_y <0 $),
without mixture of waves. Initial vortex SFH perturbations are
defined by the equations presented in the Appendix. The
calculations show a fundamental difference in the dynamics of
vortices at low and moderate $~R_{max}$.

\begin{figure}
\begin{center}
\includegraphics[width=\hsize]{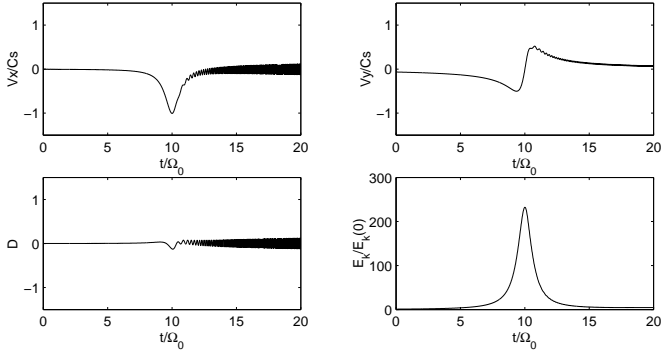}
\caption{The evolution of SFH normalized
    velocity and density perturbations ($v_x/c_s$, $~v_y/c_s$
    and $~D \equiv{\rm i} \varrho/\rho_0$) and its
    normalized energy  $~(~E_{\bf k}(t)/E_{\bf k}(0)~)~$ at
    $~R_{max}=1.06$  ($m=R/H$ in terms of the azimuthal wavenumber). In the dynamics of the initially imposed vortex
    SFH  (with $~ k_x(0)/k_y = -15$, $k_y=1$)  transient growth is dominant,
    but the wave generation starts to be  noticeable.}
\end{center}
\label{fig:lin1}
\end{figure}

\begin{figure}
\begin{center}
\includegraphics[width=\hsize]{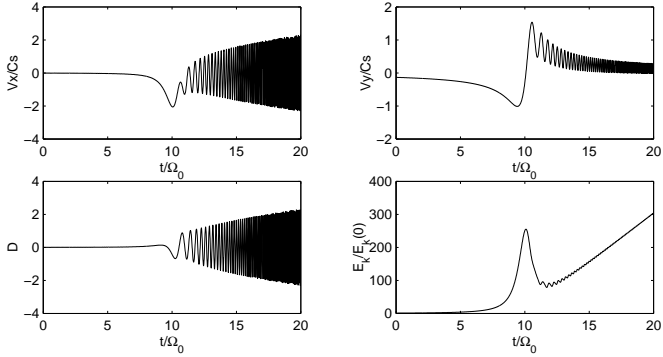}
\caption{The evolution of  SFH normalized
    velocity and density perturbations
    ($v_x/c_s$, $v_y/c_s$, $D$) and its normalized
    energy  $(~E_{\bf k}(t)/E_{\bf k}(0)~)$
    at $~R_{max}=1.34~$  ($m = 0.5 R/H$ in terms of the azimuthal wavenumber).
    Evolving in the shear flow, the initially
    imposed vortex mode SFH  (with $~ k_x(0)/k_y = -15$, $k_y=0.5$ ),
    gaining
    energy from the mean flow and amplifying, retains its
    aperiodic nature until $~k_x(t )/k_y<0$. After time
    $t^*=10$ (at which $~k_y(10)=0$), we observe the
    appearance of the wave SFH.}
\end{center}
\label{fig:lin2}
\end{figure}

\begin{figure}
\begin{center}
\includegraphics[width=\hsize]{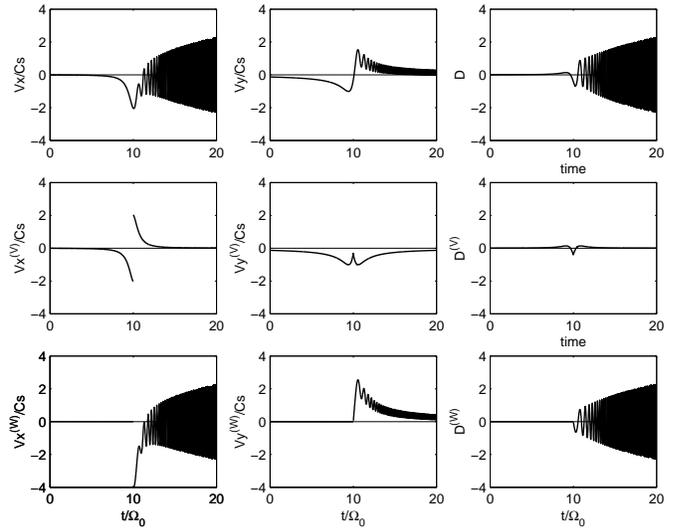}
\caption{The dynamics of $~v_x/c_s,~
v_y/c_s,~D$,
    their vortex $~(v_x^{\rm (v)}/c_s,~ v_y^{\rm (v)}/c_s,~ D^{\rm (v)})~$
    and wave parts $~(~v_x^{\rm (w)}/c_s,~ v_y^{\rm
(w)}/c_s,~ D^{\rm (w)})~$
    for a pure vortex initial perturbation ($~v_x^{\rm (w)}(0),~
    v_y^{\rm (w)}(0),~ D^{\rm (w)}(0) = 0$) with the same parameters
    as in Fig. 2. The process may be interpreted as an
    abrupt emergence of wave SFH from the related vortex SFH at
    $~t = t^*$. The amplitude of the generated wave
    $~v_y^{\rm (w)}(t^*)~$ smoothes the jump in the aperiodic
    mode (see Eq. 39).}
\end{center}
\label{fig:lin3}
\end{figure}

At $~R_{max} \ll 1$ all dynamics occurs at small $~R(t)$, at low shear
rates  compressibility is negligible and only one phenomenon occurs,
i.e. transient growth of SFH of aperiodic vortices.
The leading SFH gains energy from the mean
flow and increases its amplitude; when the SFH becomes trailing, it gives
back the energy to the flow and decreases its amplitude. It is obvious that the
transient growth of vortex perturbations exists in the
spectrally/asymptotically stable flow system (the rotation curve
derived from Kepler's law is extremely stable based on the Rayleigh
criterion). The transient growth of two dimensional vortical/aperiodic
perturbations in  an incompressible Keplerian flow was found by
Lominadze et al.  (1988). Later on, the existence of transient
growth was confirmed by Fridman (1989) and Ioannou \& Kakouris
(2001). The importance of transient growth for the onset of
hydrodynamic turbulence in Keplerian disks is discussed by
Chagelishvili et al.  (2003) and Tevzadze et al. (2003).

At moderate/large $~R_{max}$, the dynamics are richer: our
calculations show the appearance of {\bf a} conversion of vortices to
spiral-density waves at moderate shear rates. In Figs.
1-2 we present the results of
calculations at $~R_{max}=1.06;~ 1.34~$  ( respectively $ m = R/H$ and 
$m = 0.5 R/H$ in terms of the azimuthal wavenumber) . We can start to notice the conversion
at about $~R_{max}{\simeq}1~$ (i.e., at about
$~k_yH \simeq 0.7$), but the overall behavior is still dominated by
the vortex transient growth, as  is evident from the lower right
panel in Fig. \ref{fig:lin1} that shows the SFH energy growth and
decrease.  The process already becomes dominant in the SFH dynamics 
at $R_{max}>1.2~$ (i.e., at about $~k_yH<0.5$).

In Fig. 2 we present the evolution of non-dimensional perturbed
quantities $~(~v_x/c_s, v_y/c_s, D~)~$ and of the normalized energy
$~(~E_{\bf k}/E_{\bf k}(0)~)~$ of a SFH, for the case with
$~R_{max}=1.34~$. In this case, at the beginning $~R(t)\ll 1$ (as
$~k_x(0)/k_y = - 15$, i.e., $~(k_x(0)/k_y)^2 \gg 1$), and the
initial stage of the evolution is incompressible. Evolving
in the shear flow, the vortex SFH gains energy from the mean flow
and amplifies, while retaining its aperiodic nature.
$~R(t)~$ increases, becomes moderate and compressibility comes into
effect. As a result, we observe the appearance of an oscillating
part of SFH, i.e. we observe the appearance of spiral-density waves.
Thus, the linear dynamics of a vortex mode
perturbation is followed by the generation of a spiral-density wave.
When $~R(t)~$ is moderate, the time scales of the vortex and wave
SFH are comparable and the perturbations are not
separable/distinguishable, i.e. we have a mix of SFH of aperiodic
and oscillating modes.  Subsequently, $~R(t)~$ becomes small again,
the time scale of the wave SFH becomes much shorter than that of the
vortex SFH and the modes become clearly distinguishable again.

In spite of what was said above, we carry out a separation of the modes
also in the vicinity of time $~t^*$, when $~k_x(t^*)=0~$ and
$~R(t)~$ is not small. The performed separation 
 allows  us to establish the initial characteristics
of the generated wave SFH, that, in turn,  determine the further
dynamics/energetic of SFH,  even quantitatively.

In Fig. \ref{fig:lin3} we present separately the dynamics of
$v_y/c_s,$ $v_x/c_s,$ $D~$ and of their vortex $~(v_y^{\rm
(v)}/c_s,~ v_x^{\rm (v)}/c_s,~ D^{\rm (v)})~$ and wave components
$~(v_y^{\rm (w)}/c_s,~ v_x^{\rm (w)}/c_s,~ D^{\rm (w)})~$ for a
pure vortex/aperiodic initial perturbation ($~v_y^{\rm (w)}(0),~
v_x^{\rm (w)}(0),~ D^{\rm (w)}(0) = 0$), with the same parameters
used for the case in Fig. 2. From Fig. \ref{fig:lin3}
we notics:\\
-- $v_y^{\rm (v)}$ is an odd
function of the argument $~t - t^*$;\\
-- $v_x^{\rm (v)~}$ and $~
D^{\rm (v)}~$ are even functions of the argument $~t - t^*$;\\
-- At time $~t = t^*$, at which $~k_x(t^*)=0$, the vortex
mode {\it abruptly} gives rise to the corresponding wave SFH (see
Fig. 3, top panels);\\
-- At the moment of wave SFH emergence, the
phase is such that the value of $~v_x^ {\rm (w)}(t \to t^*+),~D
{\rm (w)}(t \to t^*+)=0~$ and $~v_y^ {\rm (w)}(t \to t^*+)~$ is
maximal (see Figs. \ref{fig:lin3}, bottom panels). More precisely,
\begin{equation}
|v_y^ {\rm(w)}(t \to t^*+)| =  2|v_y^ {\rm (v)}(t \to t^*-)|=2|v_y(t^*)|~;
\end{equation}
-- The excited wave harmonic (satisfying condition
$~k_x(t)/k_y>0~$) gains energy from the shear flow and is amplified; \\
-- the vortex SFH after the transient growth gives  energy back to
the background flow. \\
 This process can be altered by the action of nonlinear
forces, or by the presence of a radial entropy gradient in the
Keplerian flow, in which case vorticity is continuously generated, feeding 
the vortex with Rossby waves (see Klahr 2004).

\section{Qualitative study of Rossby waves}

In the previous section we found that the compressible shearing sheet
approximation results in a third order system of ordinary differential
equations (see Eqs. \ref{eqvx2}-\ref{eqrho2} ), or, equivalently, in a
second order inhomogeneous differential equation (see
Eq. \ref{secondorder}).
This reduction is connected to the conservation of potential vorticity and
results in the fact that one of the perturbation modes is
aperiodic/vortex. Consequently, the approximation involves two
spiral-density waves propagating in opposite directions and one vortex
mode. The vortex mode and spiral-density waves are no longer
independent: they are coupled by the strong differential character of
Keplerian rotation. As  is shown in the previous section, the vortex
mode unconditionally generates spiral-density waves, but the opposite
does not occur, i.e. the waves do not a generate vortex/aperiodic mode.

A question arises: how does the picture change when the shearing
sheet approximation is not valid, especially, when the mean vorticity
gradient across the disk is at work? We consider this problem
qualitatively, borrowing some estimates from  Tagger
(2001). At first, the linear dynamics of perturbations is
described by a third order system of ordinary differential
equations that, however, cannot be reduced to a second order
inhomogeneous differential equation, as in the previous case. In
this case all the perturbation modes are waves: two spiral-density
and one Rossby wave. The frequency of the first ones is given
by Eq. \ref{disprel}. We can get the frequency of the Rossby wave from Eq. (16)
of Tagger (2001):
\begin{equation}
~\omega_R(t) = {k_y \over {k_x^2(t)+k_y^2}} {\partial W\over
\partial r}= -{3 \over 4} {\Omega \over r} {k_y \over {k_x^2(t)+k_y^2}}~.
\label{rossby}
\end{equation}

From Eq. (\ref{rossby}) it follows that in a  thin ($H \ll r$)
Keplerian disk, for perturbations with characteristic sizes of the
order of the disk thickness ($k_x,k_y \simeq 1/H$), the Rossby wave
frequency is much less than $~\Omega$. Considering a short time
evolution ($\simeq 1/\Omega$), the disk thickness size
Rossby waves represent a non-oscillating and non-propagating vortical
perturbation. In practice, we have a degeneration of Rossby waves to
vortex mode perturbations and for them the mean flow vorticity
gradient can be neglected.

We present a simple sketch of the ``breaking of the
degeneracy'' in the wavenumber plane $(k_x,k_y)$ (see Fig. 4).
The linear dynamics of a vortical perturbation may be described by
following each of its SFH  in  the wavenumber plane. We single out
a SFH that, initially, is located at point 1 in Fig. 4,
for which $~k_x/k_y<0;~$ $k_y < 1/r;~$ $|k_x| \gg 1/r.$, i.e.
a tightly leading vortical perturbation with large
azimuthal size. Formally, the perturbation represents a Rossby
wave SFH. However, according to Eq. (\ref{rossby}), its frequency
is substantially smaller than $\Omega$. Thus,  the SFH is
aperiodic, i.e., it is degenerated to a vortex mode SFH. According
to Eq. (\ref{drift}), as $~k_x(t)$ varies in time, the SFH drifts in the
direction marked by the arrows (we present the drift of the SFH
only in the upper half-plane $~k_y>0$; since the perturbation is
real, there is a counterpart in the lower half-plane.)
Initially, as $|k_x(t)|$ decreases, the energy of the SFH grows
(see Figs. 1-3). Then, the SFH reaches the circled domain where
$|k_y|/(k_x^2+k_y^2)<r$ (point 2) and its frequency becomes no
longer negligible (the boundaries of the
domain, where breaking of degeneracy occurs, are indistinct, but
we fixed the domain to clarify the analysis). Thus, passing 
this domain, the SFH acquires a wave nature, i.e., the
degeneration to a vortex mode is broken. At the same time, the
Rossby wave SFH continues its growth that lasts until it crosses the
line $~k_x=0$  (point 3). It then generates the related SFH of a
spiral-density wave (as it does the vortex mode SFH -- see Sec. 3).
The drift after passing the line $k_x = 0$ is shown by the double line
arrows to stress the appearance of the spiral-density wave SFH and
its further drift equal to the Rossby wave drift. At point 4,
the Rossby wave SFH leaves the circled domain and becomes tightly
trailing, again degenerating into an aperiodic/vortex SFH. 
Thus, the Rossby wave appearance from a
vortex mode perturbation is not connected to a generation
process. The essence of the Rossby wave formation lies just in the
``breaking of degeneracy'' of the vortex mode perturbation. If we now choose
a vortex mode SFH with $~k_y>1/r$ (e.g., at point $1^{\prime}$), it is not
difficult to understand that the drifting SFH misses the circled
domain and the breaking of degeneration never occurs, but,
in crossing the line $~k_x=0$ (see point $3^{\prime}$), it 
generates its related spiral density wave SFH.
\begin{figure}
\begin{center}
\includegraphics[width=\hsize]{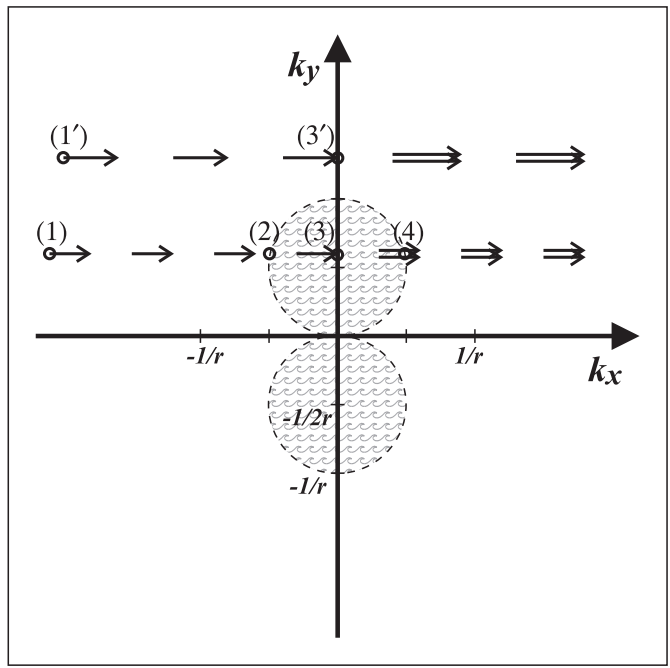}
\caption{Simple sketch of the relation between the Rossby wave and vortex
mode perturbations in the wavenumber plane ($K_x, K_y$), see details in
the text.}
\end{center}
\label{fig:Rossby}
\end{figure}

A supposition about the coupling of Rossby and spiral-density waves
is expressed in Tagger (2001): ``density waves generating Rossby
waves as they propagate, and reciprocally Rossby waves spawning
density waves as they are sheared by differential rotation''.
Chagelishvili \& Chkhetiani (1995) investigated the linear dynamics
of planetary atmospheric Rossby waves in the $~\beta$-plane
approximation, in a case of zonal flow. They found and described
the generation of spiral-density waves by Rossby waves, but not the
opposite. It may be that this asymmetry of the wave coupling is
connected to the fact that in the system, we have only one Rossby
and two spiral-density waves (propagating in opposite directions).
The same asymmetry exists in case of wave and vortex mode coupling
(see Sec. 3 and Chagelishvili et al. 1997).  In this
context it would be particularly interesting to consider the case
where a radial entropy gradient is present (Klahr 2004), since in
this case there are two independent Rossby waves. The
problem of Rossby wave generation by spiral-density waves in a given
disk model (containing both angular velocity and vorticity
gradients) will be the subject of a separate investigation.

\section{Generation of  spiral-density waves by coherent circular
vortex structure -- Numerical simulations}

\subsection{Numerical setup}

In this Section we present 2D numerical simulations of Eqs.
(\ref{hydro1}-\ref{defvgradv}) in  case of an initial coherent
circular vortex perturbation superimposed on an equilibrium
Keplerian flow.

 Since the emphasis of our numerical simulations is not only on
the vortex dynamics itself but also on the process of wave
generation, we need initial conditions corresponding to the vortex
perturbations in a pure form.

One of the ways to select the initial conditions for the vortex would be
to employ an equilibrium vortex configuration. A class of widely
used exact steady analytic solutions for single vortex patches is
given by Saffman (1992). These solutions were also used for the
construction of an elliptic vortex in Keplerian shear flow in numerical
simulations by Chavanis (2000). Another class of time-dependent vortex
solutions in uniform shear flows is given by Kida (1981). Stability
analysis of the Kida vortex revealed some aspects of its non-modal
behavior - transient growth due to the background velocity shear
(see Meacham et al. 1990). However, these vortex configurations
are solutions for incompressible flows  and cannot be directly
used in the compressible setup initially preserving the no-wave condition.

An exact equilibrium vortex solution in Keplerian shear flow is
given by Goodman et al. (1987). However, the specificity of this
solution is that it has a non-divergent velocity field, yielding a
particular enthalpy distribution to balance the compressibility
effects. In this sense, this vortex is not purely kinematic by
nature, but involves thermodynamic forces in the equilibrium
configuration.

Our aim is to perform numerical simulations of vortex dynamics for
a direct comparison and verification of results obtained within
the non-modal analysis in the wave-number space. For this purpose,
we need an initial vortex configuration which will persist in a
medium with constant sound speed - a property which is not
valid for  the equilibrium vortex derived by Goodman et al. (1987).
Hence, we provide the algorithm for the construction of a pure vortex
structure in compressible shear flow.

For simplicity we chose the initial vortex to be circular in shape
having in mind that it will be sheared into an elliptical configuration
during the evolution in the shear flow. The compressibility
effects on vortex perturbations in flows with moderate shear
parameters (as in Keplerian disks) are appreciable at $~|k_x/k_y|
\leq 1$, while the circular vortex we want to construct contains SFH
from this domain of the wave-number space. Therefore, in this case
we need a refined procedure to construct pressure, density and
velocity fields that correspond to the vortex mode perturbations in
a pure form. For this purpose we use ``pre-initial'' conditions for
the perturbations in as follows:
\begin{equation}
v_x^{\prime}(x,y,0)= \pm \epsilon_r \frac{y}{a} \left( \frac{x^2 +
y^2}{a^2} \right)^n \exp [-(x^2+y^2)/a^2]~,
\end{equation}
\begin{equation}
v_y^{\prime}(x,y,0)= \mp \epsilon_r \frac{x}{a} \left( \frac{x^2 +
y^2}{a^2} \right)^n \exp [-(x^2+y^2)/a^2]~,
\end{equation}
\begin{equation}
\rho^{\prime}(x,y,0)=0~,~~~~~~~~~~~~~~~~~~
\end{equation}
where $~\epsilon_r~$ defines the amplitude of the perturbation,
$a$  its size, $n$  the shape of the vorticity distribution and the
upper and lower signs correspond respectively to anticyclonic and
cyclonic rotation. Although the kinematics described by
the above equations seems a pure vortex, this is
not the case because the absence of the density disturbance
denotes that the equations above describe a mixture of spiral
density wave and vortical perturbations, with density disturbances
fully compensating each other. We use them to
describe a ``pre-initial'' condition, in the sense that through
them we defined the distribution of ${\cal I} (k_x,k_y)$ in
wavenumber plane. Then, on the basis of the procedure described in
the Appendix, we constructed our actual initial condition, that is
an almost pure (with minimal admixes of spiral-density waves)
circular vortex with its corresponding density field. 
The velocity field of our initial disturbance
has somewhat changed with respect to the ``pre-initial'' one.
Our initial vortical perturbation represents a rich
ensemble of leading $(k_x/k_y<0)$ and trailing $(k_x/k_y>0)$
harmonics, but only the presence of leading SFH sets the conditions for
the generation of spiral-density waves.

 We have then superimposed this perturbation on the equilibrium
Keplerian flow and have performed the simulations using our
implementation of the PPM scheme in its Eulerian version (Woodward
\& Colella 1984,  Mignone et al. 2004).  The main characteristics of
the scheme are a parabolic reconstruction that  gives third order
spatial accuracy in smooth regions of the flow,  a nonlinear Riemann
solver and time advance based on a characteristic projection that is
second order accurate. We have not used an angular momentum
 conserving form of the $\phi$ component of the momentum equation,  
however, we have tested that the  total angular momentum is 
conserved by our code 
during the simulations, with a relative accuracy of {$10^{-5}$ }.   
For long-term simulations, e.g. hundreds of orbits, one would have to use
 a scheme that conserves the local  angular momentum density (Kley 1998), but,
as we only study the evolution of 2.5 orbits and only impose a linear perturbation, 
it is possible to neglect the precise angular momentum conservation.
Since in this problem there are no natural
units of length, we have arbitrarily chosen our unit of length as
the radius at which the vortex center is located. The unit of
velocity is then chosen as the Keplerian velocity $v_{kv}$ computed
at this radius, i.e. at $r = 1$, and the disk density is taken as
units of density.  In these units the computational domain covers the
region $0.056 < r < 1.94$ and $0 < \theta < 2 \pi$, in polar
coordinates,  with a uniform (both in $r$ and in $\theta$) grid
of $1024 \times 1024$ points. We had to use such a number of grid points in 
order to have a good resolution on the vortex also covering  
a large area of the disk. In this way, for example, in the case with $a =
0.1$,  we cover the vortex with about 100 points. 
The center of the vortex perturbation
is then located at $r = 1$ and $\theta = \pi/4$. In our units, the
rotation period at the location of the vortex is therefore $T = 2 \pi$ and
our simulations reach  time $t = 2.5 T$. Boundary conditions are
periodic in the azimuthal direction and outflow in the radial
direction.   More precisely, in the radial direction we impose the zero-derivative 
of all the quantities, except the azimuthal velocity, for which we continue
the Keplerian profile and impose the zero-derivative on the perturbation. 
These conditions do not ensure perfect transmission of perturbations
impinging on the boundaries, however the portions of perturbation that are 
reflected at the  boundaries become leading waves that are deamplified by 
the shear and can influence only a small region close to the boundaries.
The main parameters of our system are then $c_s$, $a$,
$\epsilon_r$ and $n$ (see Eqs. 41-43 above).  We have considered two
values of the sound speed, two sizes of the vortex and two values
for $n$. In Table \ref{table:cases} we report the values of the
parameters used in our simulations together with some other nondimensional
parameters that may help
in comparing the importance of different effects on the vortex
dynamics. For comparison, in the  table we reported the same
parameters for the simulation performed by Godon \& Livio (1999)
and Davis (2002).

\noindent
\begin{table*}
\caption{Parameters of the simulations and estimated values of other
nondimensional parameters at the
initial vortex position. For comparison we also give values of some
parameters for Godon \& Livio (1999) and Davis (2002) simulations}
\begin{center}
\begin{tabular}{|l|c|c|c|c|c|c|c|}  \hline
    & ${c_s}/v_{kv} \left( = H/r \right)$ & $a$ & $n$ & ${\Delta d}/{r}$  &
    ${\Delta d}/{H}$  & ${v_M}/{V_{Kep}}$   &  $N$ \\
\hline A   & 0.26   &  0.1   & 0 & 0.25  & 1   & 0.0034   &  0.054       \\
\hline B   & 0.08   &  0.1   & 0 & 0.25  & 3.1 & 0.0034   & 0.054        \\
\hline C   & 0.26   &  0.01  & 0 & 0.025 & 0.1 & 0.00034  & 0.054        \\
\hline D   & 0.26   &  0.1   & 2 & 0.5   & 2   & 0.0015   & 0.015        \\
\hline Godon \& Livio 1999 & 0.15 & -  & - & 0.05 & 0.33 & 0.03 & 2.4 \\
\hline Davis 2002  & 0.075 & - & - & 0.25 & 3.3  & 0.08 & 1.3       \\
\hline
\end{tabular} \end{center}
\label {table:cases}
\end{table*}

The second column in Table \ref{table:cases} gives the value of the
sound speed (in units of the Keplerian velocity $v_{kv}$ at the
position of the center of the vortex) and, according to
Eq. (\ref{thickness}) this is equivalent to the values of the ratio
$H/r$ between the disk thickness and the radius. Case B clearly represents
the case with the lowest value of $H/r$, however all the cases
help to gain more insight into the dependence of the vortex
dynamics on the parameters: a comparison between Case A and Case B
allows us to investigate the effect of the sound speed, case C allows us to
investigate the effect of the vortex size and case D allows us to
investigate the effect of a different vorticity distribution ( a value
of $n$ different from 0 gives a vorticity ring, while $n = 0$ gives a
filled distribution).  Unlike what we said above, in case C
we made use of $1096
\times 4192$ grid points covering the domain $0.5 < r < 1.5$ and $0 <
\theta < \pi/2$ and the final integration time is $t = 3.5 $.  With this
choice we could keep the same resolution on the vortex
that we have for the other cases.  The amplitude of the perturbation
in all cases, except case C, is $\epsilon = 0.004$, and in case C it
is ten times smaller, in order to minimize the action of nonlinear
terms of Eqs. (1-4), since in this paper we want to address the linear
dynamics of the perturbation. We do this to prove the linear character
of the spiral-density wave generation process (the main aim of our
study) and to visualize the wave generation and propagation processes
clearly. In {\bf the} linear case there is no difference between the dynamics of
cyclonic and anticyclonic vortices, so, without loss of generality we
took the vortex rotation to be anticyclonic.

The parameters $\Delta d / r$ and $\Delta d /H$, as discussed in Sec.
4, giving the vortex core size $\Delta d$ in units respectively of the
radius and of the disk thickness, measure the importance of the Rossby
waves.   On the basis of Eq.
\ref{rossby} one can estimate the characteristic oscillation times of 
the Rossby waves in the simulated cases. These times significantly exceed the time
of our simulation - a few revolutions of the disk at the vortex location.
Consequently, Rossby waves do not manifest themselves in our
simulations, i.e., the real participants of the simulated dynamics are
the vortex mode and the (generated) spiral-density wave.

The parameter $N$ in the last column of Table \ref{table:cases}
measures the competition between the distortion by the shear flow (linear
phenomenon) and nonlinear phenomenona that might oppose this distorsion.
This competition may be estimated by the ratio between the vortex
maximum velocity $v_M$ and the local Keplerian velocity difference
across the vortex core radius ($ |\partial V_{Kep} / \partial r|{
\Delta d/2}$)

\begin{equation}
N = \frac{v_M}{{|\partial V_{Kep} / \partial r|}{\Delta d/2}} = 4
\frac{v_M}{V_{Kep}} \frac{r}{\Delta d}~.
\end{equation}

At $N \ll 1$ the vortex dynamics is completely linear; when $N$
becomes of the order of unity or larger, nonlinear effects are
important. As one can see from the Table, the value of this
parameter for  Godon \& Livio (1999) and Davis (2002) indicates
that in their simulations nonlinear effects are important, while
all our cases describe strongly linear dynamics of vortices.

\begin{figure*}
\begin{center}
\includegraphics[height=6cm]{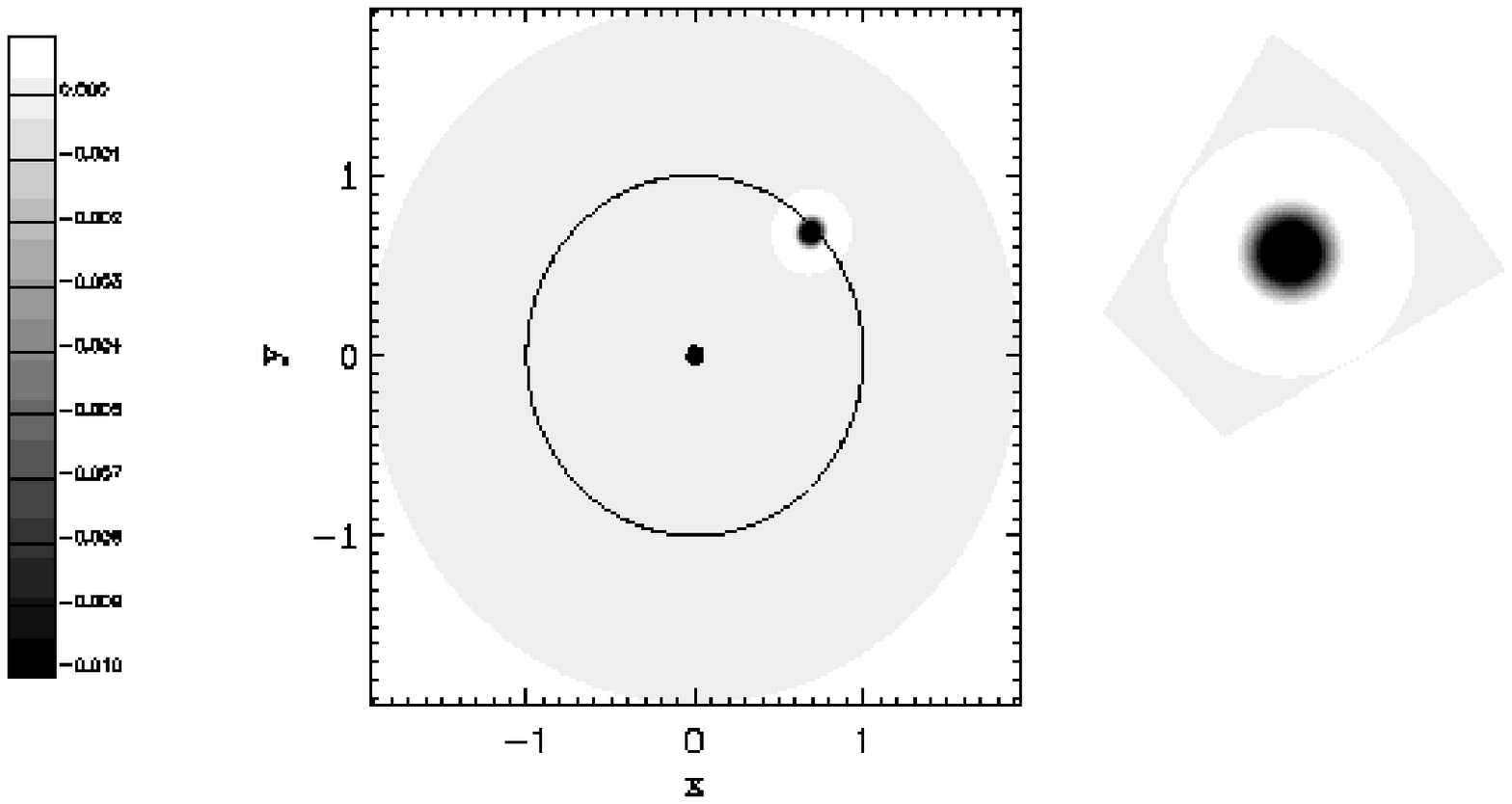}
\includegraphics[height=6cm]{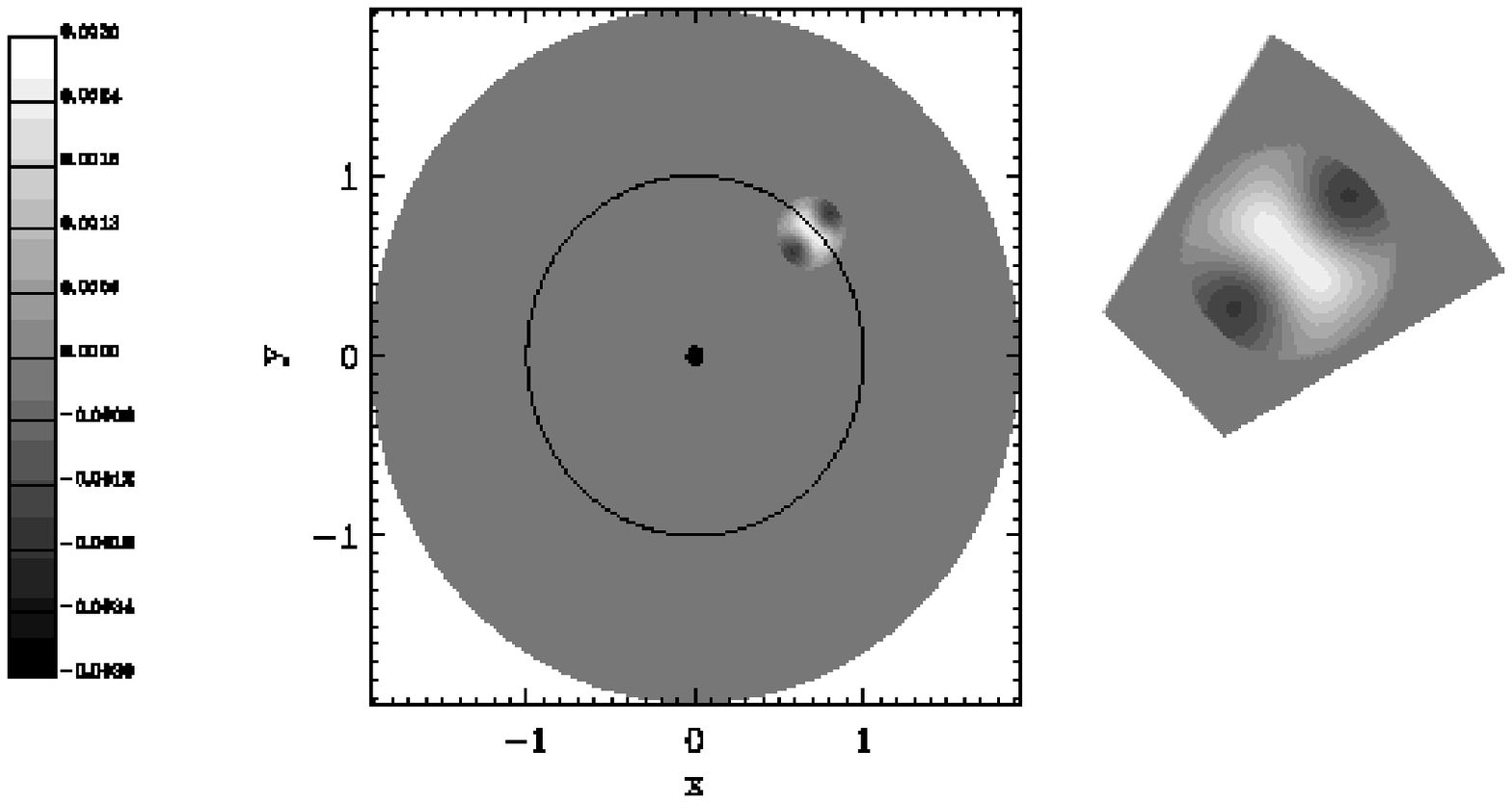}
\caption{Images of the initial density {\bf (upper plot)} and vorticity 
{\bf (lower plot)} distributions with enlargements of the vortex area.}
\end{center}
\label{fig:initsim}
\end{figure*}

\begin{figure*}
\begin{center}
\includegraphics[height=18truecm]{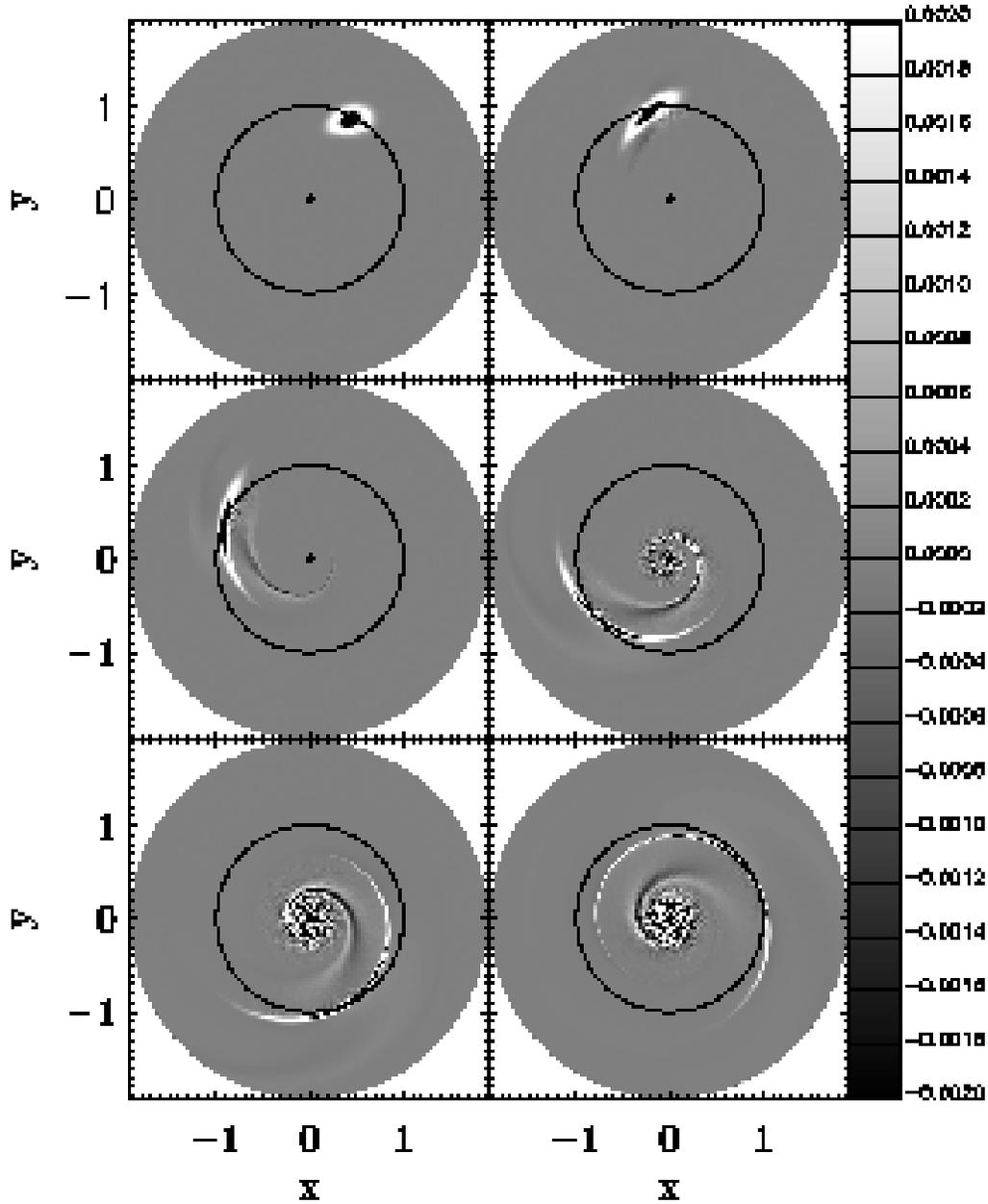}
\caption{Evolution of the vorticity distribution for case A. The six
panels show the vorticity distributions at six different times. The times are
respectively $ t = 0.31, 1, 2, 3.12, 4.5, 6.31$.  At late times the vorticity
distribution in the inner region becomes quite noisy, however one has to
take into account that the calculation of vorticity implies a 
numerical differentiation operation that amplifies the error, 
larger here due to the grid deformation }
\end{center}
\label{fig:vorticity_evol_1}
\end{figure*}

\begin{figure*}
\begin{center}
\includegraphics[height=18truecm]{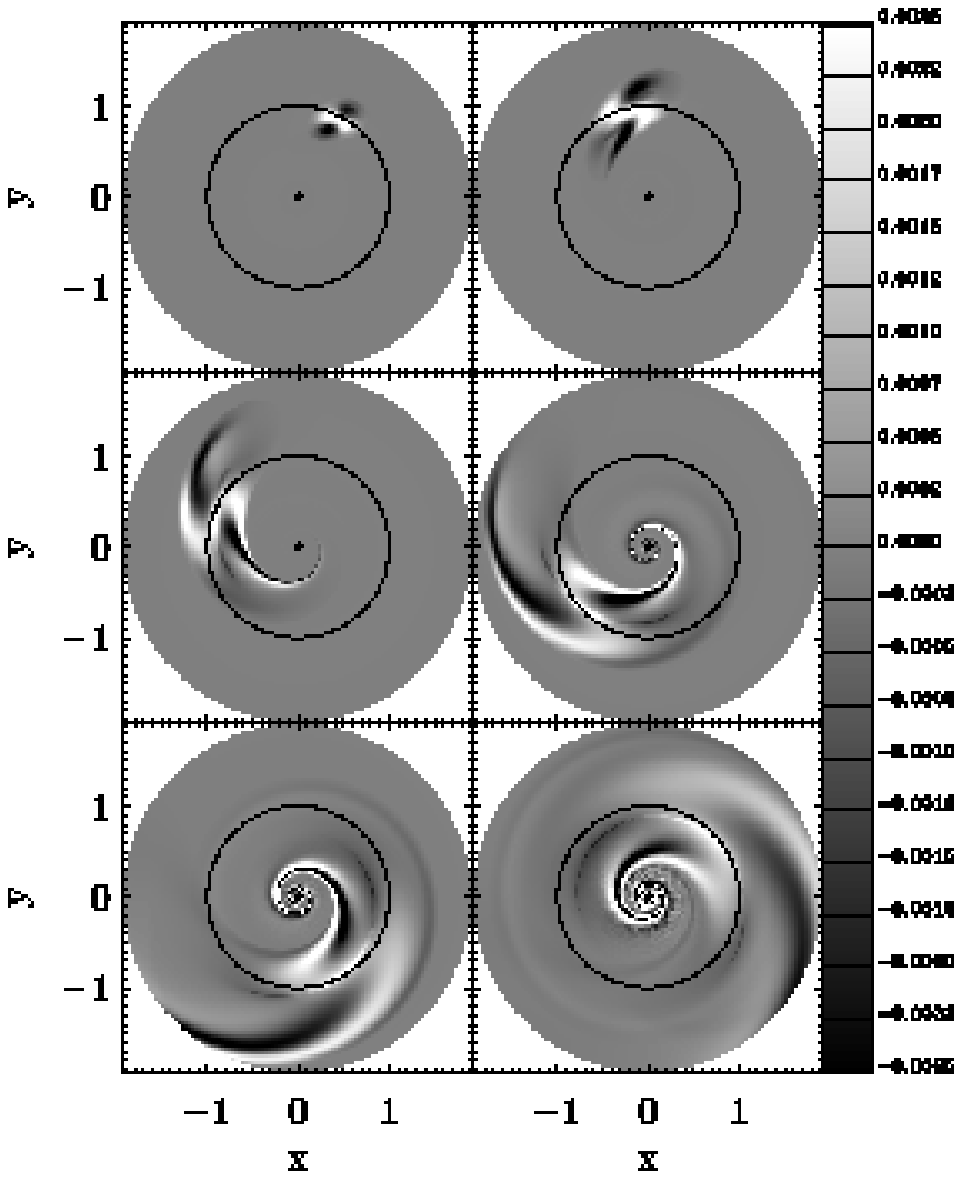}
\caption{Evolution of the density distribution for case A. The six panels show
the density distributions at six different times, $ t = 0.31, 1, 2, 3.12, 4.5, 6.31$}
\end{center}
\label{fig:density_evol_1}
\end{figure*}

\subsection{Results}

 In Fig. 5 we represent the initial conditions of our simulations (for
$n = 0$),
 more precisely we show the vorticity (upper panel) and density (lower
 panel) {\bf perturbation} fields, with enlargements of the vortex region. The vorticity
 field of the initial vortical perturbation has axial symmetry, as
 expected, while the density field has a different structure resulting
 from the balance of pressure and Coriolis forces. In Figs. 6 and 7,
 we show the initial phases of the vortex evolution for case A, up to
 $t = 2 \pi$, that corresponds to one rotation period at the initial
 vortex location. The Figures show respectively images of vorticity
 and density distributions at six different times. We see that the
 vortex is stretched by the background Keplerian flow, but we can 
 also observe that, in agreement with the theory
 presented in Sec. 3, two spiral-density waves are being generated by
 the vortex.  The stretching is well visualized in Fig.6, that shows
 the vorticity dynamics. The wave generation and propagation phenomena
 are well traced by the two arms that are visible on the images of
 Fig. 7. According to what we have shown in previous sections, a
 vortex SFH generates a wave when the SFH crosses the $k_y$ axis,
 i.e. when the wavenumber in the local radial direction becomes zero
 ($k_x = 0$, see Fig. 3 and points $3$ and $3^{\prime}$ on
 Fig. 4). Our vortex is composed of many SFH, and each of them,
 crossing the $k_y$ axis, generates the corresponding spiral-density
 wave SFH component.  Therefore, since at the moment of generation
 the wavenumber is along the azimuthal direction, we can observe the
 two generated waves to be propagating in opposite azimuthal
 directions. The waves, as they propagate, have their wavevectors
 turned by the shear flow towards the radial direction, and in time
 they are carried by the flow and become tightly trailing. The
 appearance of two arms is related to the conservation of the
 perturbation action, that initially is zero, since the vortex
 perturbation does not propagate. The total action of the generated
 wave arms should then be zero and therefore we have two arms with
 opposite propagation directions.

We now define and compute the trajectories of the excited
spiral-density waves by using the group velocity concept (of course,
trajectory, as wave group velocity, gains in importance only for
wave packages when the concept of ray is applicable). The trajectories
that we will compute will be global, but they will be based on the
local dispersion relation that can be obtained at each position in
the shearing-sheet approximation.  The validity of the
trajectories that we will compute has therefore all the
limitations of this approximation. Due to this ``half-globality''
we will indicate by $k_r(t)$ and $k_{\theta}$ (instead of $k_x(t)$ and
$k_y$ that are used in the local reference frame) the components of
the wavenumber vector respectively in the local radial and
azimuthal directions. We rewrite the
dispersion relation of spiral-density waves from Eq. (34)
\begin{multline}
\omega_s (k_r(t),k_\theta)= \pm [{\Omega}^2 +
(k_r^2(t)+k_\theta^2)c_s^2]^{1 \over 2}= \\
\pm [1+H^2(k_r^2(t)+k_\theta^2)]^{1 \over 2}c_s
\label{spiral}
\end{multline}
and then we obtain from it the radial and azimuthal components
of the group velocity
\begin{multline}
{\bf V}^G_s=\left({\partial \omega_s \over \partial
{k_r(t)}}~,~{\partial \omega_s \over \partial {k_\theta}}\right) = 
\left(\pm {k_r(t)c_s \over \sqrt{1+H^2[k_r^2(t) + k_\theta^2]}}~,~ \right. \\
\left. \pm {k_{\theta} c_s \over \sqrt{1+H^2[k_r^2(t) + k_\theta^2]}}
\right)~.
\label{vgroup}
\end{multline}

The two signs identify  the two oppositely propagating
spiral-density waves. We  notice that the group velocity is
time-dependent as $~k_r(t)~$ depends on time. We can now compute
the trajectory by defining the total wave velocity as
\begin{equation}
\frac{d {\bf r_s}}{dt} = {\bf V}^T_s={\bf V}^G_s + {\bf V}_k
\label{eq:vt}
\end{equation}
with the wavenumber that changes according to the equation
\begin{equation}
\frac{d k_r(t)}{dt}= -2A(r)k_\theta~. \label{eq:wvn}
\end{equation}
where $A(r)$ is local value of the Oort's constant. As discussed
in Sect. 3, the group velocity of spiral-density waves is directed
along the azimuthal direction at the moment of their generation,
i.e., initially we observe two oppositely propagating rays in this
direction. We can then follow the propagation by numerically integrating
Eqs. (\ref{eq:vt}-\ref{eq:wvn}). In Figs.
8 and 9 we plot the trajectories that
we have obtained in this way, superimposed on the results of the
full numerical simulations at time $t = 2$ for the two cases A
and C that differ for the initial vortex size. The different
trajectories correspond to different values of $k_\theta$ at the
emission time. The trajectories have a characteristic S-shape
since the  two rays oppositely propagating in the azimuthal
direction are progressively turned towards the radial direction
and in the meantime they are carried by the background Keplerian
flow.  We can see that our computed trajectories agree quite well
with the results of numerically simulated propagation of
spiral-density waves. We see that the agreement is better for the
case with an initially smaller vortex size. In both cases
we have part of the perturbation that is ahead of the computed
trajectory, but this part is larger for case A. This partial
disagreement can be considered as the result of the breakdown of the
approximations used in computing the trajectories and of course, the
approximations are better for a smaller size vortex.

In Fig. 10 we again plot trajectories and numerical
simulation results for case A at a later time ($t = 3.75$), to show
that the computed trajectories follow with a very good agreement
the further perturbation evolution.

Fig. 7 show a very regular structure of the emitted spiral-density
waves. {\it Prima facie}, they look like shocks, but this is not the
case. They have a {\it linear origin} and we can understand
the observed regular structure as described below. A wide spectrum
of wave SFH  is generated at each moment of time, since the
intensity of generation is different from zero over a wide range of
$~k_y$ (see Sec. 3),  and the panels in Fig. 7
represent an integral picture of the
generated and propagating wave interference. On the other hand,
according to the
graphs of $~v_x^{\rm (w)},~ v_y^{\rm (w)},~ D^{\rm (w)}~$ in Fig.
3, the wave harmonics, at the moment of generation, have very
regular phases that are similar to each other (specifically, all
density perturbation SFH are excited with zero phase -- see graph
of $~ D^{\rm (w)}~$ in Fig. 3). This phase regularity and
similarity are the roots of the regular interference picture
shown in Fig. 7.

In vortical flows, potential perturbations acquire a vortical nature and
this is the case for the spiral-density waves that we are considering.
We can estimate the
wave vorticity by Eq. (29), rewritten in  a
``half-global'' fashion (as was done defining trajectories). For
spiral density waves we have  ${\cal I}=0$:
\begin{equation} k_r(t) v_{\theta}^{\rm (w)}(t) - k_{\theta} v_r^{\rm (w)}(t)
+ 2[\Omega(r) + A(r)] D^{\rm (w)}(t) =0.
\end{equation}
Defining the wave SFH vorticity as
\begin{equation}
W^{\rm (w)} \equiv k_r(t) v_{\theta}^{\rm (w)}(t) - k_{\theta}
v_r^{\rm (w)}(t),
\end{equation}
we finally get:
\begin{equation}
W^{\rm (w)}(t) = - \frac{1}{2}\Omega(r) D^{\rm (w)}(t).
\end{equation}
From this equation we can see that the vorticity of the propagating
wave is determined by the local vorticity of the background flow and
by the amplitude of the density perturbation. Both factors grow for
the wave propagating towards the inner part of the disk and the result
is an increase of the wave vorticity in this region. For the outward
propagating wave, instead, the density amplitude still increases but
at a much lower rate than the decrease of the background flow
vorticity. Therefore the net result will be a decrease of its
vorticity.  This asymmetry is
well traced in Fig. 6, where we see a vanishing of the
 outward arms, while the inward ones become well pronounced.

\begin{figure}[tp]
\begin{center}
\includegraphics[height=8cm]{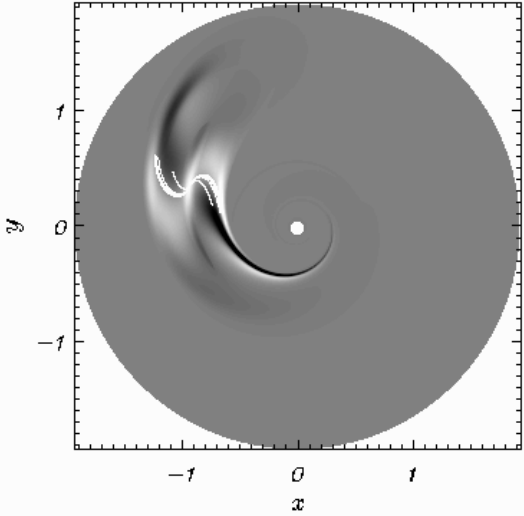}
\caption{Ray trajectories superimposed on the density distribution for
case A at time $t = 2$. The different rays correspond to different
values of $k_{\theta}$}
\end{center}
\label{fig:traja}
\end{figure}

\begin{figure}[tp]
\begin{center}
\includegraphics[height=8cm]{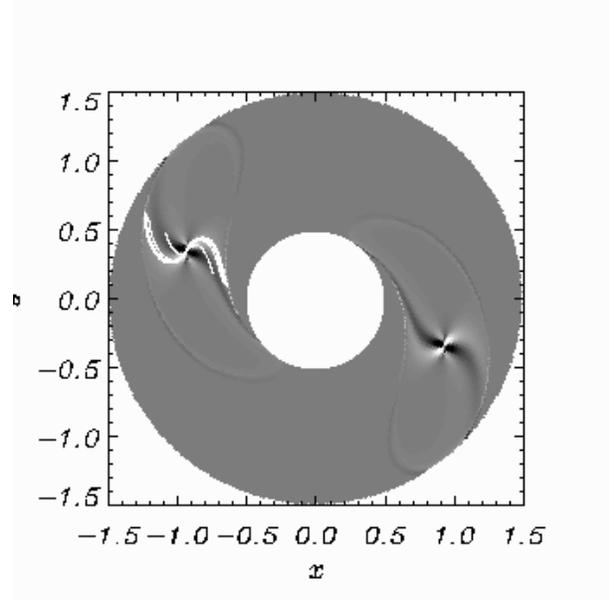}
\caption{Ray trajectories superimposed on the density distribution for
case C at time $t = 2$. The different rays correspond to different
values of $k_{\theta}$. In the image we show also a vortex in a
symmetric position for better comparison.  Note that in case C the
computational domain covers the radial interval $0.5 < r < 1.5$.}
\end{center}
\label{fig:trajb}
\end{figure}

\begin{figure}
\begin{center}
\includegraphics[height=8cm]{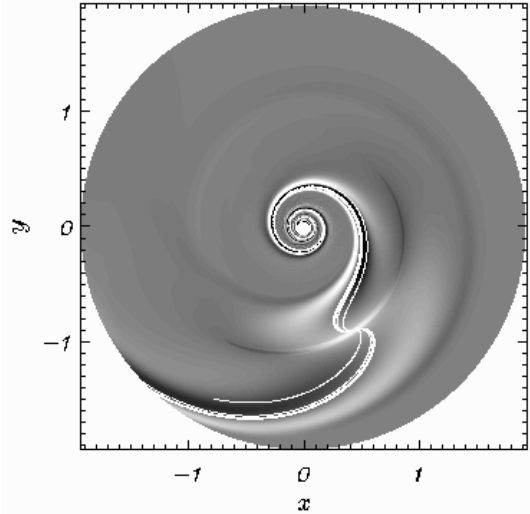}
\caption{Ray trajectories superimposed on the density distribution for
case A at time $t = 3.75$. The different rays correspond to different
values of $k_{\theta}$. }
\end{center}
\label{fig:trajc}
\end{figure}

In Fig. 11  we show grey-scale images of the density
distribution at $t=2$ for the cases A,B,C,D. Cases A and D relate
to equal sizes of parent vortex and sound speeds (see Table 1, they
differ only for the values of $n$) and
the wave propagation pictures are very similar. In
case B, due to the small value of the sound speed, the wave arms
are radially compressed and not strongly pronounced. 
In case C, due to the
small size of the parent vortex, the generated wave SFH has large
wavenumbers, their group and phase velocities are more or less the
same (close to the sound speed) and the arms are more uniform (not
diffuse).

Due to  linearity, the described phenomena have a short life time (a few
disk revolutions): in fact  vortex distortion by the
background shear flow cannot be
prevented by  velocity self-induction (a nonlinear phenomenon).
The parent vortex is stretched, becomes tightly trailing (see Fig.
6) and, in accordance with Sec. 3, is not able to generate any more
spiral-density waves. Also, the trailing vortex mode
perturbations $(k_x(t)/k_y>0)$ give back energy to the flow and
disappear (also in accordance with Sec. 3). The disappearance of
the parent vortex is evident in Fig. 12,
where we show grey-scale
images of density at later evolution times;
the three panels are respectively at $t=8.12, 11.25 ,14.5$, i.e. up to
two and a half revolutions. Spiral-density tightly trailing waves have
extended  to the whole disk region inside the vortex position, but
their generation seems to have stopped and they have disappeared from
the outer region.

 Preliminary results show that nonlinearity makes 
the parent vortex  long-lived and the wave generation process
more permanent  (Bodo et al. 2005). Indeed,
prevention of parent vortex stretching should be the result of
positive nonlinear feedback, i.e. nonlinear regeneration of the
leading SFH in the vortex spectrum. These replenished leading SFH
extract the disk flow energy and generate continuously 
spiral-density wave SFH. Consequently, they provide a permanent
generation of waves. (The concept of replenishing of SFH -- a
root of the positive nonlinear feedback -- is a theoretical concept,
numerical investigation of which is under way). 

\begin{figure*}
\begin{center}
\includegraphics[height=15cm]{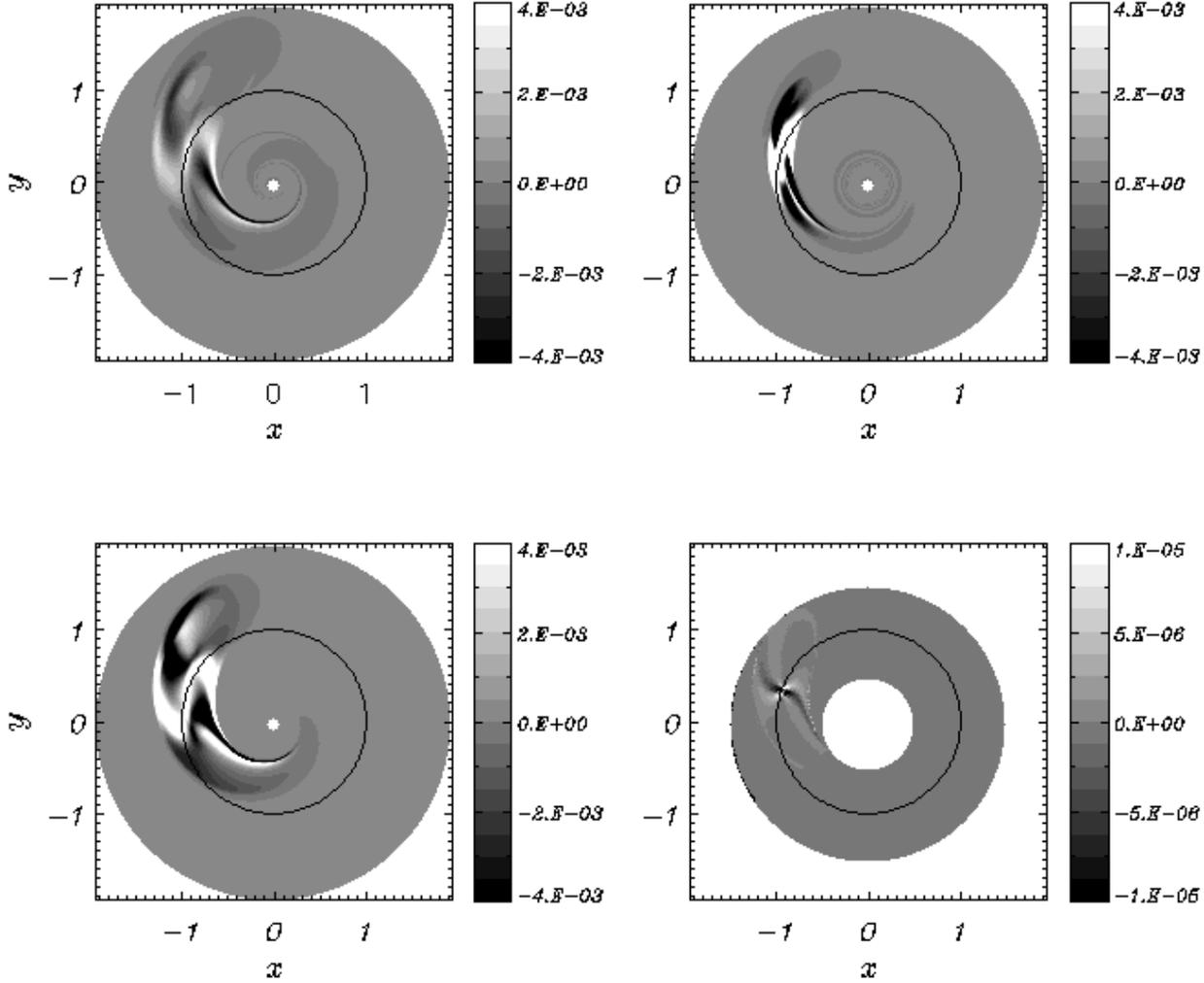}
\caption{Comparison of the four different cases at the same time. The
four panels show images of the density distribution for cases A (top left), B
(top right), C (bottom right) and D (bottom left) at $t = 2$.}
\end{center}
\label{fig:comparison}
\end{figure*}

\begin{figure}
\begin{center}
\includegraphics[height=17cm]{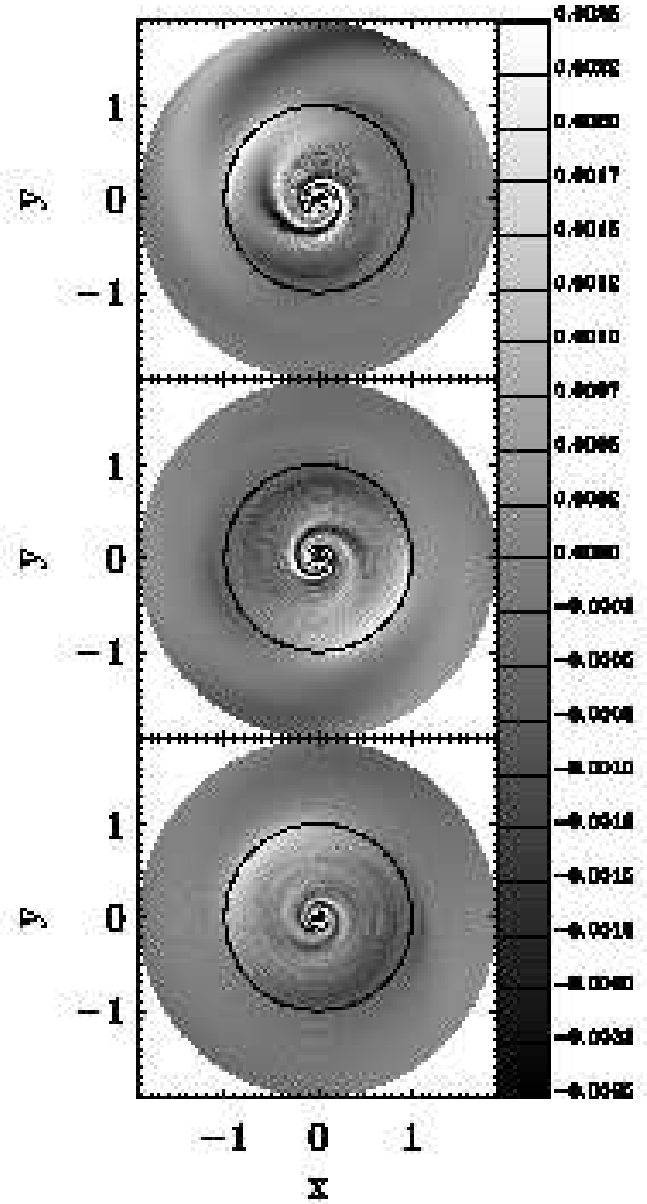}
\caption{Longer time evolution for case A. The three panels show
images of the density distributions at three different times; from top
to bottom the times are respectively $t = 8.12, 11.25, 14.5$}
\end{center}
\label{fig:rholong}
\end{figure}

The results for a single Fourier component, given in Section 3, 
show that the energy of the generated spiral density wave increases
with time, i.e. trailing spiral density waves are capable of
extracting energy from the mean flow. This wave amplification should
also be visible in our numerical calculations; for this reason, in Fig.
13, we plot the perturbation energy (as defined in Eq. 31) vs time. 

\begin{figure}
\begin{center}
\includegraphics[width=\hsize]{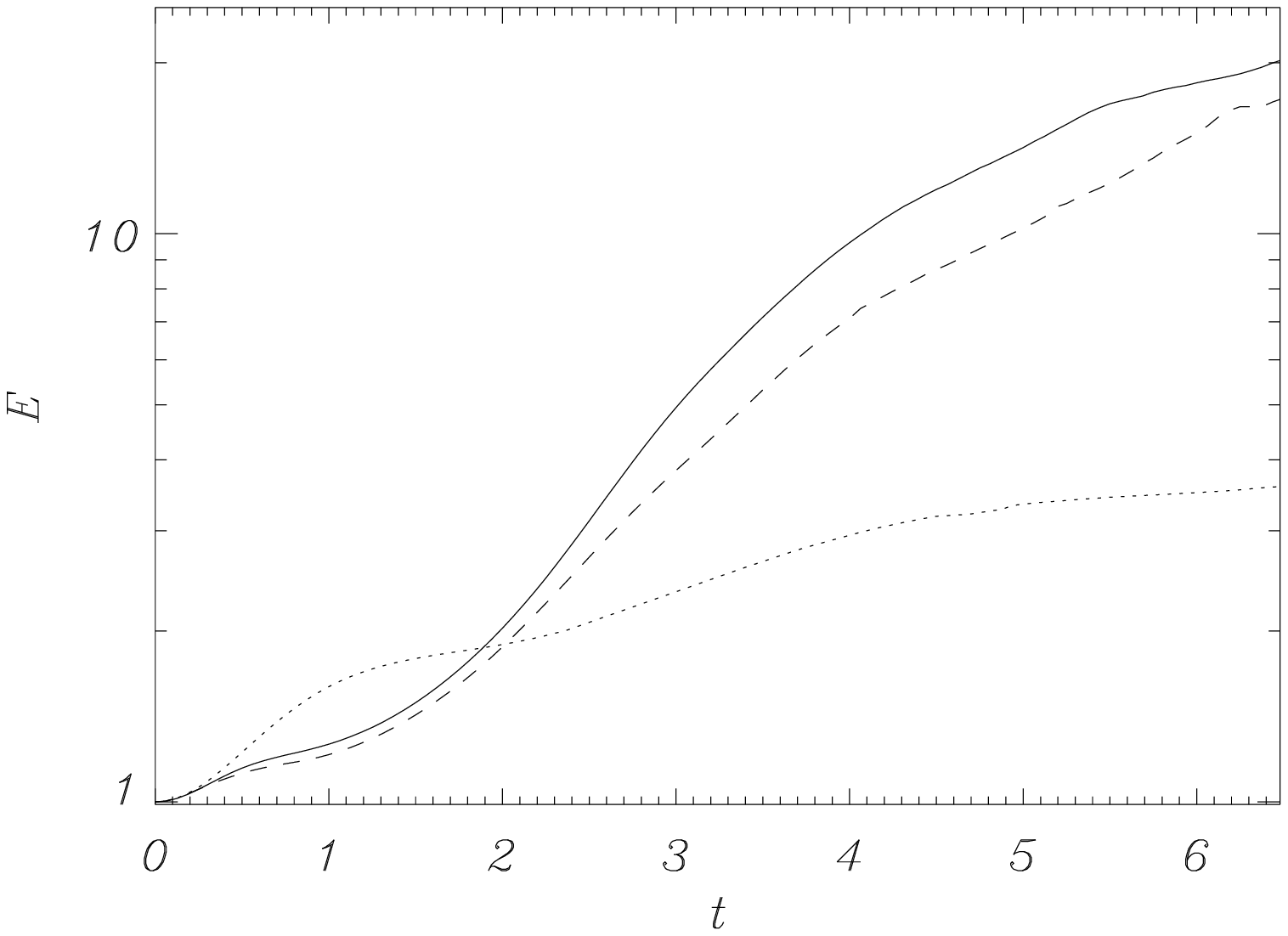}
\caption{Energy normalized to the initial value versus
time for the cases A, B and D (the solid line refers to case A,
the dotted curve to case B and the
dashed curve to case D).}
\end{center}
\label{fig:energy}
\end{figure}

 In  the figure we can see the perturbation normalized energy (normalized to
the initial value) vs. time for the cases A, B and D(case A - solid
curve,  case B - dotted curve, case D -
dashed-dotted curve). 
 The plots go up to one revolution time ($t_{max}=2 \pi$), when the
action of the parent vortex has not yet completely disappeared.
In this process of wave
generation, the vortex acts only as a mediator and the waves draw
their energy from the background flow, both at the beginning, when
they are generated, and in their propagation, when they are amplified
by the background shear flow. $k_r(t)$ of the generated wave
SFH increases with time and the dynamics of SFH becomes adiabatic,
therefore the energy of each SFH increases linearly with its
frequency: $E_{\bf k}
\sim \omega_{SD}$.  Generally, according to Eq. \ref{disprel}
\begin{equation}
\omega_{SD}^2={ \Omega^2 +
[k_r^2(t)+k_{\theta}^2]c_s^2},
\label{eq:Omega-k}
\end{equation}
but, at moderate sound speeds (cases A and D), in Eq.
(\ref{eq:Omega-k}) the second term becomes quite rapidly dominant and,
as time passes, one can write $E_{\bf k} \sim k_r(t)c_s$, i.e. the
energy of each SFH increases linearly with time.  
The increase of $k_r$ is determined by the local shear
rate, that increases in the inner region, at smaller values of the
radius $r$. It is easily understood that at low sound speeds (case
B) in Eq. (\ref{eq:Omega-k}) the first term is dominant for a long
time and moreover the increase in $k_r$ is slower, since the waves
propagate more slowly towards the interior and they therefore experience a
smaller shear rate. Consequently, $\omega_{SD}$ and $E_{\bf k}$ are
practically constant.

\begin{figure}
\begin{center}
\includegraphics[width=\hsize]{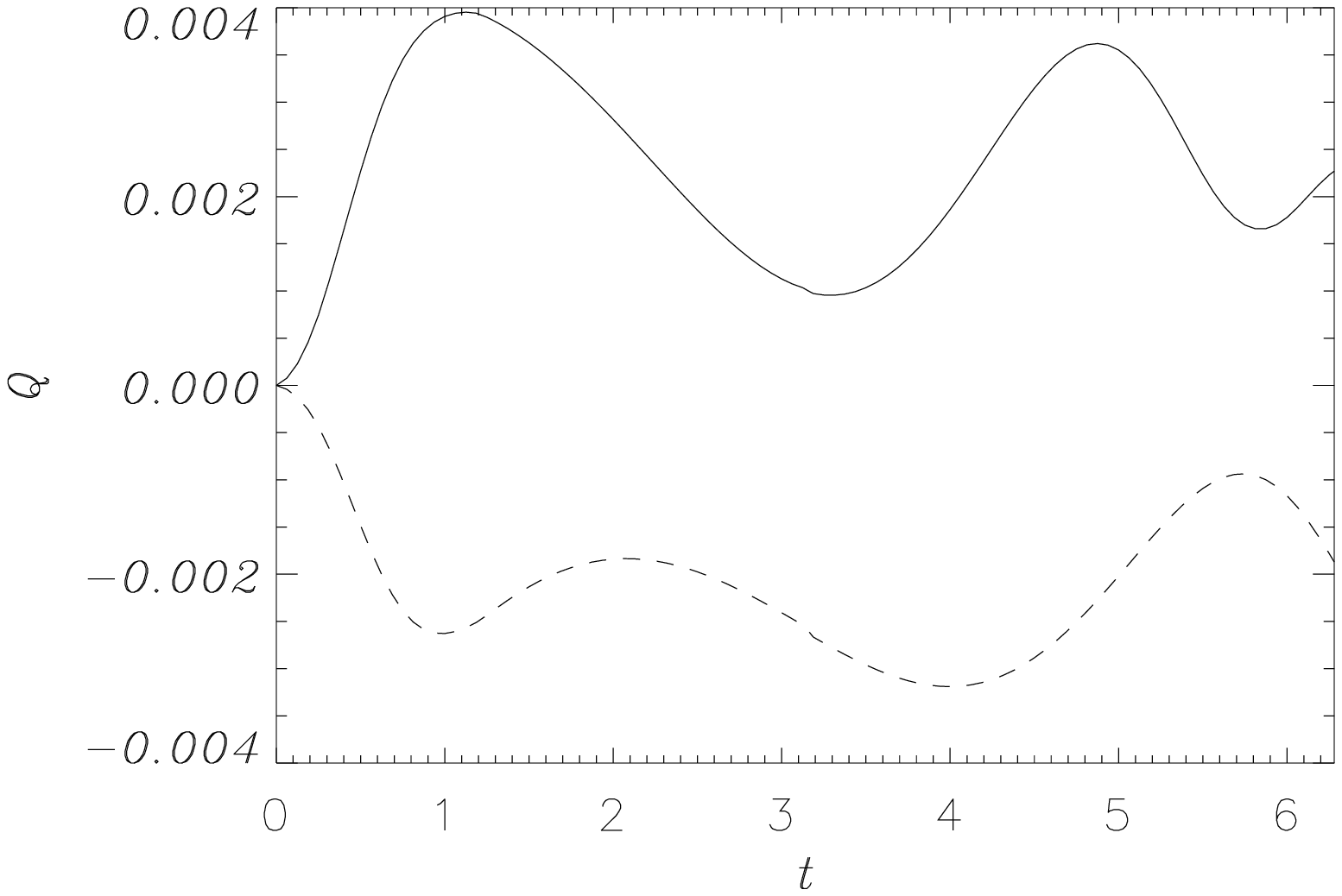}
\caption{Plot of the the total mass that has flown
through radius $r$ at time $t$ versus time, for case A. This figure refers to
radial positions that are immediately inside ($r = 0.875$ , solid line) and
immediately outside ($r = 1.125$, dashed line) the vortex.}
\end{center}
\label{fig:mflux1}
\end{figure}

\begin{figure}
\begin{center}
\includegraphics[width=\hsize]{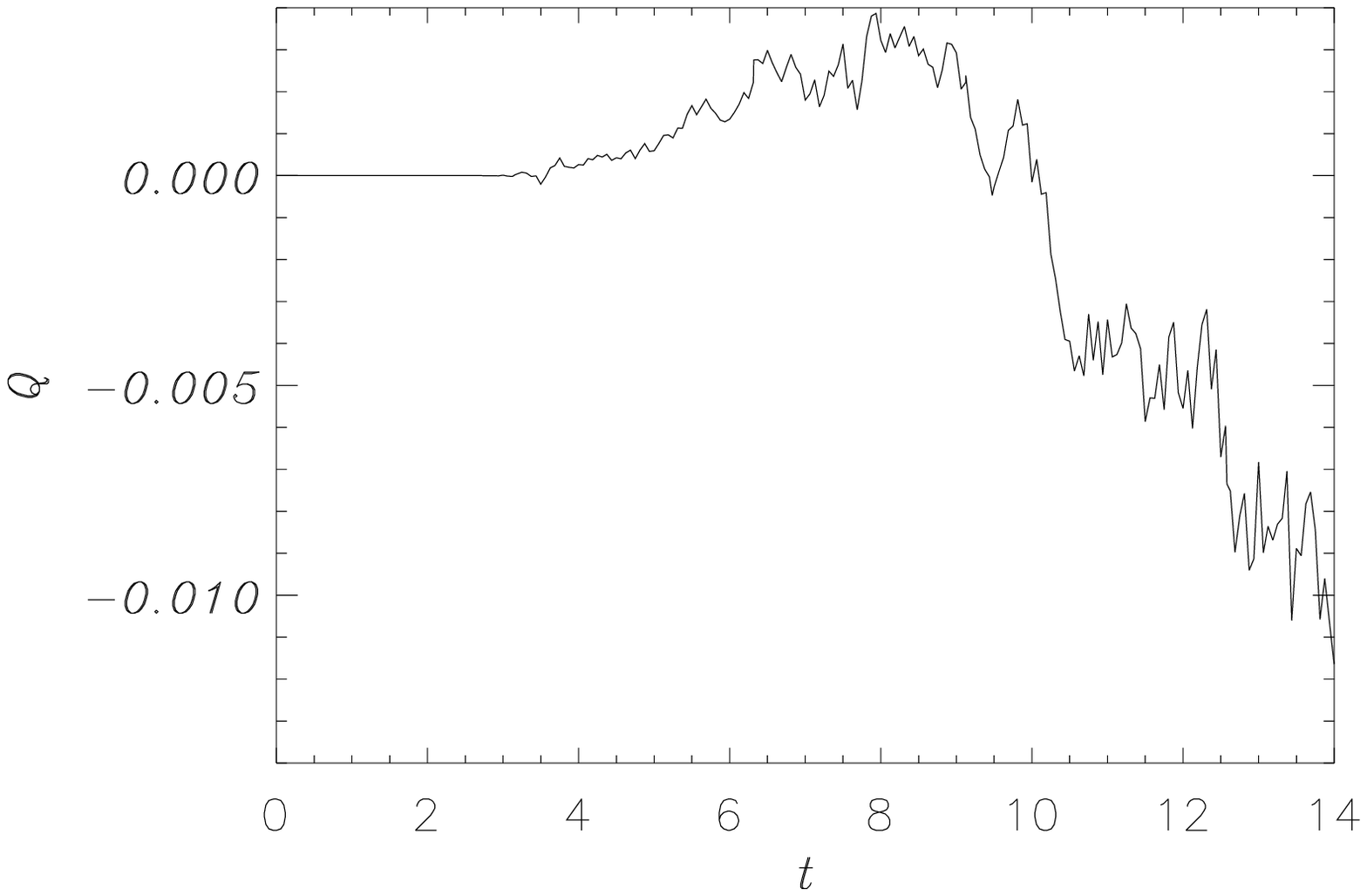}
\caption{otal mass that has flown
through radius $r$ at time $t$ versus time, for case A. This figure refers to
$r = 0.1$.}
\end{center}
\label{fig:mflux3}
\end{figure}

A feature of the generated spiral density waves that we have
  discussed above is their regularity, i.e. perturbations associated
  with the waves have a  prominent sign and do not average to
  zero. One consequence of this is that, in principle,
  it may be possible to have a mass flux associated with them, 
  that would be  therefore proportional to the wave amplitude.
  We have tested this possibility by plotting the function:
\begin{equation}
Q (r,t) = \frac{1}{2\phi}\int_{0}^{t}{\rm d} t^{\prime}
\int_{0}^{2\pi} {\rm d}{\theta} \rho v_r(r,\theta,t^{\prime})
\end{equation}
as a function of time, at different radial positions ($r = 0.875$ and $r =
1.125$ in Fig. 14, $r = 0.1$ in Fig. 15). This
function $Q(r, t)$ represents the total mass that has flown
through radius $r$ at time $t$, in units of the mass contained inside
the initial vortex radial position.  Fig. 14 represents this quantity
for positions that are immediately inside ($r = 0.875$ , dashed line) and
immediately outside ($r = 1.125$, solid line) the vortex. The figure shows
that the action of the waves seems to be in the direction of
mass collecting  towards the vortex area. Fig. 15  represents,
instead, the mass flow induced by the waves in the inner disk
region (at $r = 0.1$). Here we see a definite tendency of the wave to
induce a non-negligible  inflow of mass. These plots
suggest  a general trend of the phenomenon. and  more
realistic results with simulation of longer time dynamics of
large amplitude coherent vortices are needed (Bodo et al. 2005).
As discussed above, since we are
essentially considering a linear process and since the quantity ${\cal
M}$ is a first order quantity, the values that  we get are
proportional to the initial amplitude of the perturbation we give. The
case that we present is an anticyclonic vortex; simulations of a
cyclonic vortex show that all the fluxes are reversed, as they should
be. Of course, in addition to this mass flux that is proportional to the
perturbation amplitude, the trailing spiral density waves induces an
outward angular momentum transfer that is quadratic in the
perturbation amplitude and, therefore, negligible in these linear calculations.

\section{Discussion and conclusions}

In this paper we present a linear, non-resonant phenomenon of
spiral-density wave generation by vortices in a Keplerian disk
flow that is closely related to the non-normality of linear dynamics of
perturbations in smooth (without inflection points) shear flows.
This phenomenon can be well interpreted by the use of the non-modal
approach, i.e. by following in time the linear dynamics of spatial
Fourier harmonics of the vortex mode perturbations. We have done this
in the shearing sheet model and we found that spiral-density wave
are generated only by the
leading vortex mode SFH that meet the condition $~k_x(0)/k_y<0$.
The generation takes place at relatively large wavelengths of SFH
-- it becomes noticeable at about $~k_yH \simeq 0.7$ and it is
dominant in the SFH dynamics at $~k_yH<0.5$.  At large
wavenumbers (small wavelength) all dynamics of SFH runs at low
shear rates and only one phenomenon occurs: transient growth of
SFH of aperiodic vortices.  At small
wavenumbers (large wavelength) when $~k_yH<0.5$, the dynamics in
addition to the transient growth involves conversion of vortices
to spiral-density waves. After their appearance, the energy of the
wave SFH increases and the wave SFH itself in time becomes  tightly
trailing $~k_x(t)/k_y \gg 1$.

Small amplitude (linear) circular vortex structures (that
represent a rich ensemble of leading and trailing SFH) generate
two oppositely propagating wave arms in accordance with the perturbation
action conservation law. They have very regular structure and
{\it prima facie} look like shocks, but this is not the case. They have
a linear origin and are the result of the interference of the
generated spiral-density waves. Their outward and inward propagating
arms are asymmetric due to the disk curved geometry and flow
asymmetry ($\Omega(r)$ increases outward and decreases inward).

Since spiral-density wave generation is governed by linear forces,
it does not depend on the sign of the vorticity of initial
aperiodic perturbation and should be equal for
cyclonic and anticyclonic vortices even in the nonlinear regime.
Davis (2002) (whose simulations are in a nonlinear regime - see Table
2), reported that a coherent cyclonic vortex
simultaneously emits a wave as  does an anticyclonic one
(nonlinear cyclonic and anticyclonic vortices differ
by their lifetime -- cyclonic vortices are destroyed in
appreciably shorter times than anticyclonic ones). This result
 may be considered as  circumstantial evidence for the
 linear origin of the wave.

The fact that the waves are generated by a linear mechanism (and
not nonlinear ones, as was thought ) appreciably increases
the significance of waves in the disk matter inflow/outflow
processes and we showed that waves generated by anticyclonic vortices
tend to collect matter in the vortex region, which may be important for
protoplanetary disks, and tend to induce  inflow of matter
in the inner disk regions.

The described dynamics  are short-lived due to the small amplitude of
the parent vortex. One can speculate that the parent vortex could 
be long-lived and
the dynamical picture more permanent at larger amplitudes, when
nonlinear self-induction of the vortex velocity prevents its shear
distortion. Indeed, nonlinear regeneration of the leading SFH
($(k_x/k_y<0)$) in the vortex spectrum can prevent the  vortex from 
stretching. These replenished leading
SFH gain the disk flow energy and continuously generate spiral-density wave SFH. 
Thus, permanent nonlinear replenishment of leading
SFH of the parent vortex could generate permanent disk flow energy
 extraction and  wave  generation.  The effective 
efficiency of the feedback mechanism in permanently maintaing the 
vortex and overcoming the linear and nonlinear damping effects inherent
to a Keplerian disk (Hawley, Balbus and Winters 1999) however has still
to be proven by numerical simulations and analytical means.

The linear phenomenon of spiral-density wave generation by vortices
investigated in this paper is quite universal; it occurs in disk flows
with high shear rates
and should be seen in different kind of astrophysical disks
(protoplanetary disks; quasars; thin and thick galactic and binary
system disks).

\begin{acknowledgements}
This work is supported by ISTC grant  G-553 and GRDF grant 3315 and
by MIUR grant 2002028843. G.D.C. and A.G.T. would like to
acknowledge the hospitality of Osservatorio Asrtonomico di Torino.
Numerical calculations were partly performed in CINECA
(Bologna, Italy) thanks to the support by INAF.
The authors would like to thank H. Klahr for helpful comments that improved 
the paper.
\end{acknowledgements}

\newpage

\section{Appendix - Selection of Initial Values of Vortex Mode Perturbations}

The initial values of physical quantities for our two dimensional
direct numerical simulations of Eqs. (1-4) may be composed by the
superposition of the background flow and local vortex mode
perturbation centered at some $r_0$:
$$
{\bf V}(r,\phi,0) = {\bf
V_0}(r,\phi,0) + {\bf v^\prime}(x,y,0) ,
\eqno(A.1a)
$$
$$
P(r,\phi,0) = P_0 + p^\prime(x,y,0) ,
\eqno(A.1b)
$$
$$
\rho(r,\phi,0) = \rho_0 + \rho^\prime(x,y,0) .
\eqno(A.1c)
$$
where $x$ and $y$ is defined by
$$
x \equiv r - r_0 ; ~~~~~ y
\equiv r_0 (\phi - \Omega_0 t) ; ~~~~~{x \over r_0}, {y \over r_0}
\ll 1  .
\eqno(A.2)
$$

The background is the Keplerian flow (${\bf V_0} =
(0,V_{0\phi},0), ~V_{0\phi}^2({\bf r}) = {G M / r}$) with
homogeneous pressure and density $(P_0,~\rho_0 = const)$. The
rigorous selection of initial vortex mode perturbation in the
(compressible) flow requires a refined procedure, as vortex
(aperiodic) modes acquire nonzero divergence -- nonzero
density perturbation (see discussion in Sec. 3). In order to
select initial vortex mode perturbations we use the initial
value problem in the wavenumber plane ({\bf k}-plane):
\begin{multline}
 \left(
\begin{array}{c} v_x^{\prime}(x,y,0) \\ v_y^{\prime}(x,y,0) \\ \rho^\prime(x,y,0)\\
p^\prime(x,y,0)
\end{array}
\right) = \int \int {\rm d}k_x {\rm d}k_y \left(
\begin{array}{l}
v_x(k_x,k_y,0) \\ v_y(k_x,k_y,0) \\ \varrho (k_x,k_y,0)\\
p(k_x,k_y,0)
\end{array}
\right) \times \\
\times \exp({i} k_x x + {i} k_y y) .
\tag{A.3}
$$
\end{multline}

Rewriting Eqs. (29,31):
$$
{{d^2 v_y(t)} \over {d t^2}} +
[4\Omega_0(\Omega_0 + A) + k^2(t) c_s^2 ]v_y(t)= k_x(t)c_{\rm s}^2
{\cal I} ,
\eqno(A.4)
$$
$$
k_x(t) v_y(t) - k_y v_x(t) + 2(\Omega_0
+ A) D(t) =const \equiv {\cal I} .
\eqno(A.5)
$$
Hence, $~{\cal
I}$, $~v_y(0)~$ and $~ \left[ {{d v_y(t)} / {d t}} \right]_{t=0}~$
form the full set of initial conditions for Eq. (A.4). We seek
the vortex mode solution in the following analytic form:
$$
v_y(0)
\equiv v_0(0) + v_1(0) + ~...~ + v_n(0) + ~...~ ,
\eqno(A.6)
$$
where the zero order term is deduced from the stationary form of
the solution (see Eq. A.4 at $~A=0~$) and subsequent terms are
derived using the iterative method (see also Chagelishvili et al.
1997):
$$
v_0(0) \equiv [v_0(t)]_{t=0} \equiv \left[{k_x(t) \over
{\cal K}}\right]_{t=0} {\cal I}(k_x,k_y),
\eqno(A.7)
$$
$$
v_n(0) \equiv [v_n(t)]_{t=0} \equiv - \left[{1\over {\cal
K} c_s^2}{{d^2 v_{n-1}(t)}~ \over {d t^2}}\right]_{t=0} ,
\eqno(A.8)
$$
where
$$
{{\cal K} \equiv {k_H^2+k_y^2+k_x^2}} ,
\eqno(A.9)
$$
$$
k_H^2 \equiv {{4\Omega_0(\Omega_0 + A)} \over
{c_{s}^2}} .
\eqno(A.10)
$$

The principal concern with such a solution is its convergence. For
instance, this solution diverges at $~t=t^*$, when $~k_x(t^*)=0~$
$(~A \not= 0~)$. The area of the divergence strongly depends on
the value of the shear rate. In our analysis we consider moderate
shear rates, when $~A/(k_yc_s) \leq 1$. In this case the above
series is convergent for times when $~|k_x(t)/k_y|>1~$. Moreover,
the first two term of this series appears to be an excellent
approximation to the exact numerical solution for times when
$~|k_x(t)/k_y| \geq 2~$. Therefore, we can use this analytic
solution to compose the vortex mode perturbations that are
localized in the wavenumber plane outside the $~|k_x(t)/k_y| \leq
1~$ area. We calculate the $v_y(k_x,k_y,0)$ and it's derivative
for the vortex mode perturbations in explicit form:
$$
v_y(k_x,k_y,0) = \biggl( {k_x \over {\cal K}} + {8A^2 k_xk_y^2
\over c_s^2} \cdot {3k_H^2 + 3 k_y^2 - k_x^2  \over{\cal K}^4}
\biggr) {\cal I}(k_x,k_y) ,
\eqno(A.11)
$$
\begin{multline}
v_y^\prime(k_x,k_y,0) = \\
\biggl( 2A k_y {{k_x^2 - k_y^2 - k_H^2 }
\over {\cal K}^2} - {16 A^3 k_y^3 \over c_s^2 } \times \\
\times { 3k_H^4 +6k_H^2k_y^2 -
24k_H^2k_x^2 + 3 k_y^4 + 5k_x^4 - 24 k_x^2 k_y^2 \over {\cal K}^5
} \biggr) \times \\
\times {\cal I}(k_x,k_y) ,
\tag{A.12}
\end{multline}
where
$$
v_y^\prime(k_x,k_y,0) = \left[ {{\rm d} \over {\rm d} t}
v_y(k_x(t),k_y,t) \right]_{t=0} .
$$
Remaining physical quantities
may be defined using the following equations:
$$ v_x(k_x,k_y,0) =
{1 \over {4(\Omega_0 + A)^2 + k_y^2 c_s^2}} \biggl[ k_x k_y c_s^2
v_y(k_x,k_y,0) - ~~~~~~~~~~~~~~~~~
$$
$$ - 2 (\Omega_0 +
A)v_y^\prime(k_x,k_y,0) + k_y c_s^2{\cal I}(k_x,k_y,) \biggr] ,
\eqno(A.13)
$$
$$ {p(k_x,k_y,0)\over c_s^2}=\varrho(k_x,k_y,0)
=~~~~~~~~~~~~~~~~~~~~~~~~~~~~~~~~~~~~~~~~~~~~~~~~~~~~
$$
$$= {{\rm
i} \rho_0 \over 2 (\Omega_0 + A)} \biggl[ k_x v_y(k_x,k_y,0) - k_y
v_x(k_x,k_y,0) - {\cal I}(k_x,k_y) \biggr] .
\eqno(A.14)
$$

Hence, using any distribution of the potential vorticity $~{\cal
I}(k_x,k_y)~$ (or $~v_y(k_x,k_y,0)$)  and Eqs. (A.9-14) we can
construct corresponding initial vortex mode perturbation in the
wavenumber plane.

\end{document}